\documentclass[aps,prl,onecolumn,showpacs]{revtex4}

\usepackage{graphicx}
\usepackage{dcolumn}
\usepackage{amsmath}
\begin{document}

\title{Compensated electron and hole pockets in an underdoped high $T_{\rm c}$ superconductor}

\author{Suchitra~E.~Sebastian$^1$}
\author{N.~Harrison$^2$}
\author{P.~A.~Goddard$^3$}
\author{M.~M.~Altarawneh$^2$}
\author{C.~H.~Mielke$^2$}
\author{Ruixing Liang$^{4,5}$}
\author{D.~A.~Bonn$^{4,5}$}
\author{W.~N.~Hardy$^{4,5}$}
\author{O.~K.~Andersen$^{6}$}
\author{G.~G.~Lonzarich$^1$}

\affiliation{
$^1$Cavendish Laboratory, Cambridge University, JJ Thomson Avenue, Cambridge CB3~OHE, U.K\\
$^2$National High Magnetic Field Laboratory, LANL, Los Alamos, NM 87545\\
$^3$Clarendon Laboratory, Department of Physics, University of Oxford, Oxford OX1~3PU, UK
$^4$Department of Physics and Astronomy, University of British Columbia, Vancouver V6T 1Z4, Canada\\
$^5$Canadian Institute for Advanced Research, Toronto M5G 1Z8, Canada
$^6$Max-Planck-Institut fuer Feskoerperforschung, Stuttgart, Germany
}
\date{\today}

\begin{abstract}
We report quantum oscillations in the underdoped high temperature superconductor YBa$_2$Cu$_3$O$_{6+x}$ over a wide range in magnetic field 28~$\leq\mu_0H\leq$~85~T corresponding to $\approx$~12 oscillations, enabling the Fermi surface topology to be mapped to high resolution. As earlier reported by Sebastian {\it et al.} [Nature {\bf 454}, 200 (2008)], we find a Fermi surface comprising multiple pockets, as revealed by the additional distinct quantum oscillation frequencies and harmonics reported in this work.  We find the originally reported broad low frequency Fourier peak at $\approx$~535~T to be clearly resolved into three separate peaks at $\approx$~460~T, $\approx$~532~T and $\approx$~602~T, in reasonable agreement with the reported frequencies of Audouard {\it et al.}~[Phys. Rev. Lett {\bf 103}, 157003 (2009)]. However, our increased resolution and angle-resolved measurements identify these frequencies to originate from two similarly-sized pockets with greatly contrasting degrees of interlayer corrugation. The spectrally dominant frequency originates from a pocket (denoted $\alpha$) that is almost ideally two-dimensional in form (exhibiting negligible interlayer corrugation). In contrast, the newly resolved weaker adjacent spectral features originate from a deeply corrugated pocket (denoted $\gamma$). On comparison with band structure, the d-wave symmetry of the interlayer dispersion locates the minimally corrugated $\alpha$ pocket at the `nodal'  point ${\bf k}_{\rm nodal}=(\pi/2,\pi/2)$, and the significantly corrugated $\gamma$ pocket at the `antinodal' point ${\bf k}_{\rm antinodal}=(\pi,0)$ within the Brillouin zone. The differently corrugated pockets at different locations indicate creation by translational symmetry breaking $-$ a spin density wave is suggested from the strong suppression of Zeeman splitting for the spectrally dominant pocket, additional evidence for which is provided from the harmonics we resolve in the present experiments. In a broken-translational symmetry scenario, symmetry points to the nodal ($\alpha$) pocket corresponding to holes, with the weaker antinodal ($\gamma$) pocket corresponding to electrons $-$ likely responsible for the negative Hall coefficient reported by LeBoeuf {\it et al.} [Nature {\bf 450}, 533 (2007)]. Given the similarity in $\alpha$ and $\gamma$ pocket volumes, their opposite carrier type and the previous report of a diverging effective mass in Sebastian {\it et al.} [Proc. Nat. Am. Soc. {\bf 107}, 6175 (2010)], we discuss the possibility of a secondary Fermi surface instability at low dopings of the excitonic insulator type, associated with the metal-insulator quantum critical point. Its potential involvement in the enhancement of superconducting transition temperatures is also discussed.
\end{abstract}
\maketitle

\section{ I. Introduction}
Different bosonic physics from tightly bound pairs of fermions may arise depending on whether the binding takes place between like
particles or between particles and holes~\cite{kohn1}. Such bosonic physics can potentially be involved in the physics of high temperature
superconductivity. On one hand, strongly interacting pairs of like particles with zero momentum have been proposed to constitute the fabric of unconventional superconductors~\cite{micnas1,anderson1}, while on the other, electron-hole pairs with finite momentum could condense into a competing 
state with a superlattice, such as a spin density wave. Another realisation of such particle-hole pairing could occur in the limit of strong Coulomb coupling, where electron and hole pockets of identical size become susceptible to an excitonic insulator instability~\cite{jerome1,halperin1,allender1}. In this study, we use high resolution quantum oscillation measurements to show that the underdoped high temperature superconductor
YBa$_2$Cu$_3$O$_{6+x}$~\cite{liang1} contains compensated electron and hole pockets, potentially predisposing it to excitonic electron-hole pair condensation.

Quantum oscillations have been measured on underdoped YBa$_2$Cu$_3$O$_{\rm 6+x}$ using in-plane transport~\cite{doiron1,leboeuf1}, out-of-plane transport~\cite{ramshaw1}, torque~\cite{sebastian2,sebastian3,jaudet1,audouard1}, and contactless conductivity~\cite{sebastian1,sebastian4,singleton1} experiments. In this work, the power of quantum oscillations is enhanced by the technique we employ that utilises interlayer dispersion measurements to enable a momentum-space identification of individual Fermi surface sections$-$ details of which are provided in Section II. Quantum oscillations are measured on high quality detwinned single crystals of YBa$_2$Cu$_3$O$_{\rm 6+x}$ (x=0.54, 0.56)~\cite{liang1} as a function of magnetic field and of angle. Measurements are made down to $\approx$ 1~K  with the sample immersed in $^4$He medium, and the contactless conductivity technique used to obtain a high value of signal-to-noise ratio. Out-of-plane rotation measurements are performed in pulsed magnetic fields up to 65~T, and two-axis (in-plane and out-of-plane) rotation measurements are performed in DC fields up to 45~T. Additional experimental and analysis details are provided in Appendix A.

\begin{figure}[ht!]
\begin{center}
\includegraphics*[width=.65\textwidth]{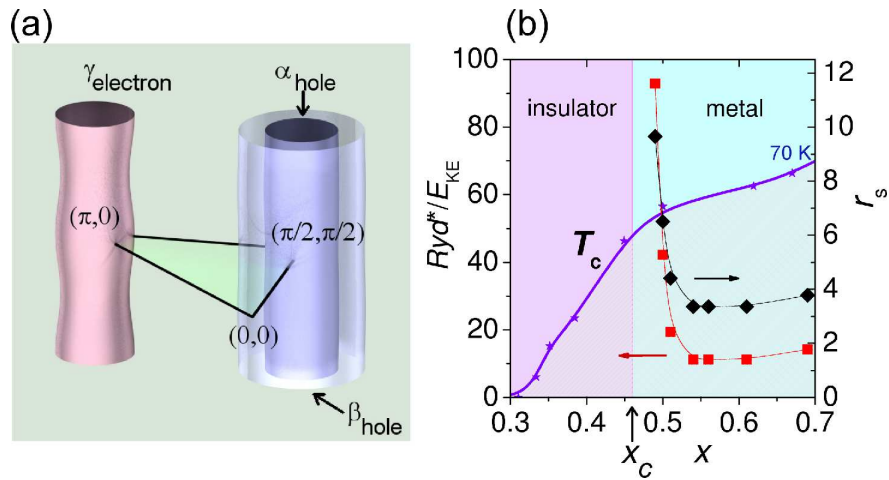}
\caption{{\bf a}, A schematic of the small hole (blue) pocket $\alpha$ and small electron (red) pocket $\gamma$, illustrating greatly contrasting
degrees of corrugation and approximate locations within the sector of the Brillouin zone bounded by $(0,0)$, $(\pi/2,\pi/2)$ and $(\pi,0)$, shown
for $-2\pi<k_z<2\pi$. Also shown in semi-transparent fashion is the larger hole orbit $\beta$ (corresponding to a separate Fermi surface section or potentially a magnetic breakdown orbit~\cite{sebastian2}). The degree of corrugation is determined from
angle-dependent Fourier transforms shown in Figs.~\ref{oscillations} and frequencies shown in Fig.~\ref{angles}a. On using the Onsager relation,
the three pockets collectively yield an effective hole doping of $p=$~11.7~$\pm$~0.2~\%. {\bf b}, Estimates of the ratio of the exciton binding
energy to kinetic energy $Ryd^\ast/E_{\rm ke}$ and effective exciton separation $r_{\rm s}$ as a function of oxygen concentration $x$, using the expressions in Appendix B and effective masses from Refs.~\cite{sebastian1,singleton1}, with arrows indicating the appropriate
axes. Also shown are the corresponding values of $T_{\rm c}$ (renormalized) from Ref.~\cite{liang1} and different shadings representing the
insulating and metallic regimes at low $T$.} \label{pockets}
\end{center}
\end{figure}

Our angle-resolved measurements show that the dominant series of low frequency oscillations~\cite{doiron1} arises not from one, but from two pockets of similar size, the distinct topologies of which point to different positions in the Brillouin zone on comparison with the interlayer dispersion, which has d-wave symmetry in underdoped YBa$_2$Cu$_3$O$_{6+x}$~\cite{andersen1,pavarini1} (see Fig.~\ref{pockets}a). 
On comparison with the interlayer hopping integrals in YBa$_2$Cu$_3$O$_{6+x}$~\cite{andersen1,pavarini1}, we show that the prominent frequency corresponding to the $\alpha$ pocket with minimal interlayer corrugation is likely to represent a section of Fermi surface at the $(\pm\pi/2,\pm\pi/2)$ nodal location in the Brillouin zone where holes are expected to nucleate on doping. In contrast, the more recently measured satellite frequencies~\cite{audouard1} of smaller amplitude correspond to a strongly corrugated pocket ($\gamma$) close to the antinodal location in the Brillouin zone, suggestive of electron carriers.

\section{II. Results}
\subsection{Interlayer corrugation}
\begin{figure}[ht!]
\begin{center}
\includegraphics*[width=.4\textwidth]{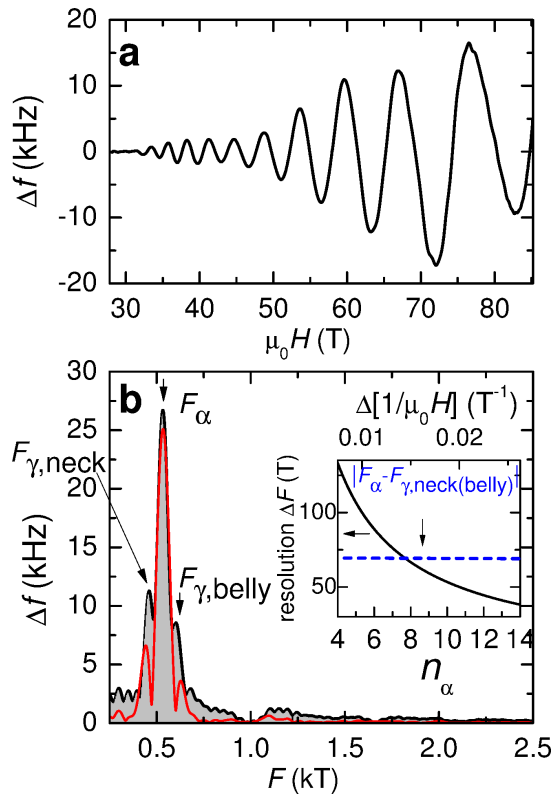}
\caption{{\bf a} Quantum oscillations made using the contactless conductivity method on a sample of YBa$_2$Cu$_3$O$_{6.54}$ over a broad range of magnetic fields 28~$<\mu_0H<$~85~T at $\approx$~1.5~K, corresponding to $\approx$~12 oscillations of the prominent $\alpha$ frequency. The experimental data below and above 65~T are from measurements made using the same sample on the same probe but in two different pulsed magnets. {\bf b}
A Fourier transform of the quantum oscillations both without (black curve) and with (red curve) a Hann window applied. The inset shows a comparison of the separation in frequency $F_{\gamma,{\rm neck}}-F_\alpha$ or $F_\alpha-F_{\gamma,{\rm belly}}$ with the frequency resolution limit $\Delta F_{\rm lim}\approx[1/\Delta(1/\mu_0H)]$ (left axis) and with the equivalent number of $\alpha$ oscillation periods $n_\alpha$ (bottom axis)$-$ the frequency separation exceeds the frequency resolution for $n_\alpha\gtrsim$~8, enabling all three frequencies to be clearly resolved in Fourier transforms and fits of the quantum oscillations.} \label{resolution}
\end{center}
\end{figure}

High experimental resolution of the quantum oscillation frequency is key to resolving and characterising the two adjacent peaks flanking the larger central peak (shown in Fig.~\ref{resolution}). The weaker spectral features at 460~$\pm$~2~T and 603~$\pm$~2~T on either side of the central frequency at 533~$\pm$~2~T are distinguished on account of their separation being well above the frequency resolution limit of our current experiment ($\Delta F_{\rm lim}\approx[1/\Delta(1/\mu_0H)]\approx$~42~T). The three distinct peaks in the Fourier transform (shown in Fig.~\ref{resolution}) are observed both without (black curve) and with (red curve) a Hann window to reduce diffraction artefacts.

\begin{figure*}[ht!]
\begin{center}
\includegraphics*[width=.65\textwidth]{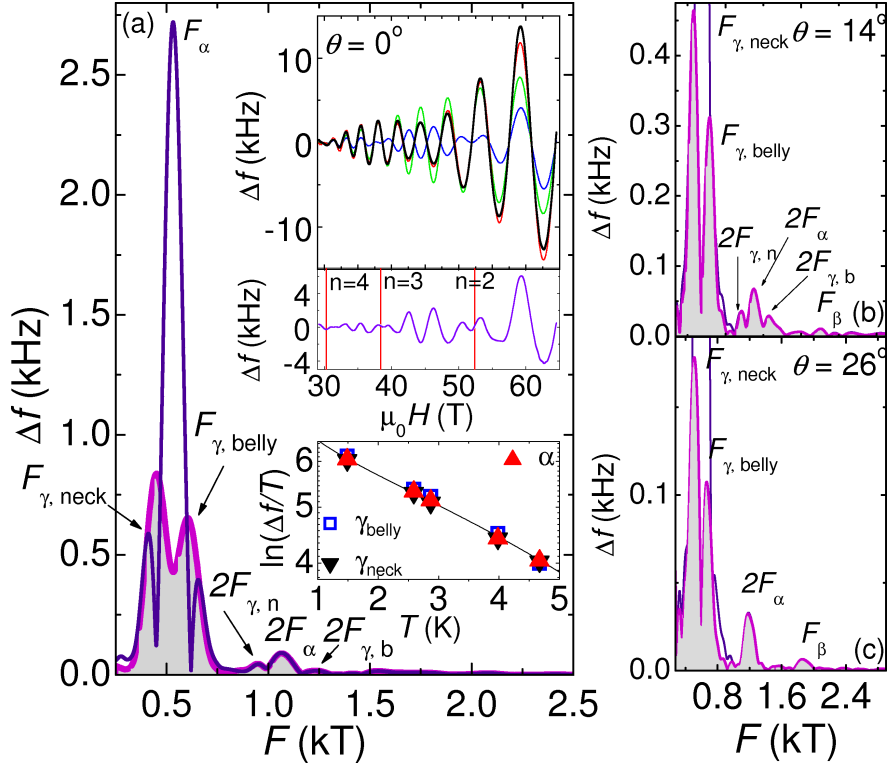}
\caption{{\bf a}, Fourier transform before (purple) and after (magenta) subtracting the $\alpha$ component of a fit to Eqn.~\ref{fittwo} (green line in the upper inset) to the quantum oscillations (black line in the upper inset) measured in a sample of YBa$_2$Cu$_3$O$_{6+x}$ with $x=$~0.56 at $T\approx$~1.5~K using the
contactless conductivity technique (see Appendix A). A polynomial is used to fit the background and the waveform is modulated by a Hann
window prior to Fourier transformation. The prominent oscillations and harmonics are indicated. Extremal Fermi
surface cross-sections given by $A_k=(2\pi e/\hbar)F$ are obtained from the observed frequencies $F_\alpha$, $F_{\gamma,{\rm neck}}$ and
$F_{\gamma,{\rm belly}}$ and harmonics $2F_\alpha$, $2F_{\gamma,{\rm neck}}$ and $2F_{\gamma,{\rm belly}}$ as indicated; here $H=|{\bf H}|$. The red line in the upper inset is an unconstrained fit to Eqn.~\ref{fittwo}, where the green and blue lines are the $\alpha$ and $\gamma$ components respectively. The purple curve (upper inset, lower panel) shows the result of subtracting the $\alpha$ component (green line) of the unconstrained fit to Eqn.~\ref{fittwo} from the raw data, revealing a distinctive beat pattern in the lower panel of the inset. The vertical dotted lines represent the fields at which nodes occur for $n=2,3,4$ from Eqn.~\ref{nodeequation} using the fit value of $\Delta F_{\gamma,0}$. The lower inset shows the temperature-dependences
of the $F_\alpha$, $F_{\gamma,{\rm neck}}$ and $F_{\gamma,{\rm belly}}$ amplitudes (the latter two renormalized to equalize the amplitudes at the
highest temperature), indicating similar effective masses $m^\ast$ for the $\alpha$ and $\gamma$ pockets. A fit to $\Delta
f=a_0R_T$~\cite{shoenberg1} yields $m^\ast_\alpha$=1.56~$\pm$~0.05~$m_{\rm e}$, $m^\ast_{\gamma,{\rm
neck}}=m^\ast_{\gamma,{\rm belly}}=$~1.6~$\pm$~0.2~$m_{\rm e}$ (where $m_{\rm e}$ is the free electron
mass). {\bf b} \& {\bf c}, Examples of Fourier transforms of oscillations measured in a sample with $x=$~0.54 at $T\approx$~1.5~K and different
angles $\theta$ (data in Figs.~\ref{angles}b and c) before and after subtracting the fitted $\alpha$ component (see Fig.~\ref{angles}b) as described in
the text (expanded to show the finer high frequency structure)$-$ further Fourier transforms before and after subtraction are shown in the Appendix D. Note that the higher frequency $F_\beta$ (when observed) is 3.14~$\pm$~0.05 times larger than $F_\alpha$, inconsistent with a harmonic of $F_\alpha$~\cite{sebastian2,sebastian1}.
} \label{oscillations}
\end{center}
\end{figure*}

\begin{figure*}[ht!]
\begin{center}
\centerline{\includegraphics*[width=.8\textwidth]{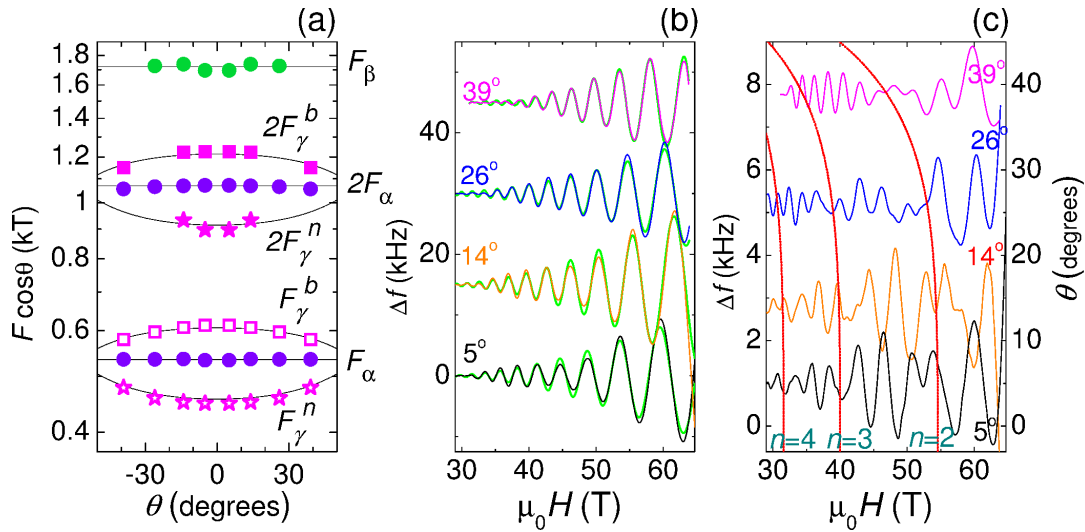}} 
\caption{{\bf a}, The product $F\cos\theta$ (for both fundamental and harmonic~-~solid symbols for unsubtracted data and hollow symbols for subtracted data)
obtained from the Fourier analysis shown in Fig.~\ref{oscillations}b and c and Appendix D plotted versus $\theta$. $F_{\gamma,{\rm neck}}$ and $F_{\gamma,{\rm belly}}$ are obtained by Fourier analysis after subtracting the fitted $\alpha$ component (using Eqn.~\ref{fittwo}). 
For clarity, the data have been symmetrized with respect to $\pm\theta$: both the swept $\theta$ data (shown in
Fig.~\ref{twoaxis}a) and measurements made over extensive range of $\theta$ in Ref.~\cite{sebastian3} indicate the oscillations to be symmetric
in $\pm\theta$, as expected for a detwinned single-phase orthorhombic crystal. 
Lines indicate the expected $\theta$ dependences according to Eqns.~\ref{neckbelly} and \ref{yamaji}, using the fitted parameters in Fig.~\ref{oscillations}a. Consistency with Eqn.~\ref{yamaji} indicates a deeply
corrugated $\gamma$ section of Fermi surface~\cite{wosnitza1,yamaji1}. The harmonics are obtained directly from peaks in Fourier transforms of the raw data$-$ consistency between fundamentals and harmonics confirms our finding of similarly-sized but distinctly different $\alpha$ and $\gamma$ Fermi surface sections. {\bf b}, Examples of quantum oscillations measured in YBa$_2$Cu$_3$O$_{6+x}$ with $x=$~0.54 at different angles $\theta$, as indicated (where $\phi\approx$~0) together with the fitted $\alpha$ component (green lines, using Eqn~\ref{fittwo}) as described in the text. {\bf c}, The quantum oscillations measured in YBa$_2$Cu$_3$O$_{6+x}$ with $x=$~0.54 at different angles $\theta$ after subtracting the green fits in ({\bf b}), yielding residuals dominated by the $\gamma$ pocket oscillations. Curves are offset according to $\theta$ (right axis). Also plotted (red lines) are the predicted values of $H_n$ from Eqns.~\ref{nodeequation} and \ref{yamaji} using the fit value of $\Delta F_{\gamma,0}$.} \label{angles}
\end{center}
\end{figure*}

Evidence for the three resolved low frequency oscillations originating from similar volume electron and hole pockets with different degrees of corrugation (shown schematically in Fig.~\ref{pockets}a) in addition to the previously resolved high frequency oscillation~\cite{sebastian2} is presented in Fig.~\ref{oscillations}, which shows quantum oscillations and Fourier transforms
measured at different angles $\theta$ between the crystalline $c$ axis and ${\bf H}$, and in Fig.~\ref{angles} where the separately resolved
quantum oscillation frequencies ($F_\alpha$, $F_\beta$, $F_{\gamma,{\rm neck}}$ and $F_{\gamma,{\rm belly}}$) are plotted versus $\theta$. The subtle beat pattern modulating the amplitude of the measured oscillations (Fig.~\ref{oscillations}a, upper panel of upper inset) and the multiple low frequency peaks in the Fourier transform suggest two different Fermi surface sections yielding closely spaced low frequencies. To separate $\alpha$ and $\gamma$ pocket contributions, we use three independent analysis methods and find them to yield mutually consistent results for a significant difference in corrugation between the $\alpha$ and $\gamma$ pockets.

Given the layered character of the cuprate family of high temperature superconductors, the quasi-two dimensional nature of the Fermi surface needs to be factored into the quantum oscillation analysis. Specifically, for quasi-two dimensional materials, interlayer corrugation introduces phase smearing that leads to a distinctive beat pattern which has been well characterised~\cite{wosnitza1}. The location and separation in magnetic field of the amplitude zeroes (nodes) constituting the beat pattern are a topologically constrained function of the depth of corrugation, and may be used to extract the size of the corrugation. For this purpose, we use an expression that explicitly treats the form of phase smearing in layered systems~\cite{mineev1} (Eqn.~\ref{fittwo}) rather than the standard Lifshitz-Kosevich expression~\cite{shoenberg1} which does not capture this form of interference (see Appendix G).

We perform an unconstrained fit (red line) of two corrugated Fermi surface sections to the measured oscillations in the 65~T magnet (black line) for a sample of oxygen concentration $x=$~0.56 at $\theta=0$ (Fig.~\ref{oscillations}a, upper inset). Oscillations are fit to the expression
\begin{equation}\label{fittwo}
\Delta f=\sum_{i=\alpha,\gamma}\Delta f_{i,0}R_TR_{\rm D}{\rm J}_0\big(\frac{2\pi\Delta F_{i,\theta}}{\mu_0H\cos\theta}\big)\cos\big(\frac{2\pi
F_{i}}{\mu_0H}\big)
\end{equation}
incorporating a summation over independent $\alpha$ and $\gamma$ corrugated cylinders~\cite{wosnitza1,mineev1,harrison1}. The spin splitting factor~$R_{\rm s}$~\cite{shoenberg1} does not appear in this expression since it has been shown to be close to unity in previous measurements (ref.~\cite{sebastian3} and Appendix C). Here, $\Delta f_{i,0}$ is the amplitude, $R_T = (2\pi^2k_{\rm B} m^\ast_i T/e\hbar\mu_0H)/\sinh(2\pi^2k_{\rm B} m^\ast_i T/e\hbar\mu_0H)$ is a temperature damping factor where $m^\ast_i$ is the quasiparticle effective mass obtained from the fit shown in Fig.~\ref{oscillations}, $R_{\rm D} = \exp\big(\frac{-\Gamma_{i}}{\mu_0H}\big)$ is a Dingle damping factor where $\Gamma_i$ is a damping constant~\cite{shoenberg1}, $\Delta F_{i,\theta}$ is the depth of corrugation of sheet `$i$' at angle $\theta$, while the Bessel function ${\rm J}_0$ captures the interference due to phase smearing, with characteristic `neck' and `belly' extremal frequencies 
\begin{equation}\label{neckbelly}
F_{i,{\rm
neck,belly}}\cos\theta=F_{i}\pm\Delta F_{i,\theta}
\end{equation}
for each section~\cite{wosnitza1,mineev1,harrison1}. 
On performing an unconstrained fit to Eqn.~\ref{fittwo} we obtain corrugations $\Delta F_{\alpha,0}\lesssim~11$~T and $\Delta F_{\gamma,0}=~72\pm 4$~T, while $F_{\alpha,0}=~531\pm 2$~T and $F_{\gamma,0}=~532\pm 2$~T are the same within experimental uncertainty. These parameters correspond to three frequencies $F_{\gamma,{\rm neck}}=$~460~$\pm$~4~T, $F_{\alpha,0}=$~531~$\pm$~2~T (the splitting being unresolvably small) and $F_{\gamma,{\rm belly}}=$~604~$\pm$~4~T in close agreement with those resolved in the Fourier transform in Fig.~\ref{resolution}. Similar frequencies were recently reported in Ref.~\cite{audouard1}, in which case the experimental uncertainty was larger due to the reduced range in $1/H$ (see Fig.~\ref{resolution}b inset). The analysis presented here is performed on data spanning between $\approx$~9 and 12 measured oscillations.

The monotonic field dependence of the $\alpha$ component (green line) identified in the present experiment implies that it can be reliably subtracted from the raw data (on performing a fit to Eqn.~\ref{fittwo}), enabling the residual beat pattern of the $\gamma$ oscillations (purple line) to be extracted (lower panel of the upper inset to Fig.~\ref{oscillations}a). Distinct nodes are observed in the residual $\gamma$ oscillations. We turn to the expected locations of zeros in the amplitude due to the interference pattern to confirm that they coincide with the observed nodes. The values of magnetic field at which amplitude zeros are expected due to interference are given by
\begin{equation}\label{nodeequation}
\mu_0H_{n,i}\cos\theta=\frac{8\Delta F_{i,\theta}}{4n+3}.
\end{equation}
Eqn.~\ref{fittwo}~\cite{mineev1} can be expanded as the interference between neck and belly frequencies $F_i-\Delta F_{i,\theta}$ and $F_i+\Delta F_{i,\theta}$ respectively. The neck and belly frequencies interfere destructively whenever they have a relative phase difference equal to an odd multiple of $\pi$, which occurs at values of the field $H_n$ given by Equation~\ref{nodeequation}.

Using the fit value of $\Delta F_{\gamma,0}$, the expected nodes for $n=2,3,4$ are indicated by the vertical red lines in the lower panel of the upper inset to Fig.~\ref{oscillations}a. We find these to coincide very well with the observed nodes in the residual $\gamma$ oscillations, establishing that the frequencies $F_{\gamma,{\rm neck}}\approx$~460~T and $F_{\gamma,{\rm belly}}\approx 604$~T correspond to the neck and belly extremal frequencies of a single corrugated $\gamma$ cylinder, distinct from the $\alpha$ section. Indeed, these frequencies correspond to those that appear in the extracted fourier transform in Fig.~\ref{oscillations} (depicted before and after subtraction of the fitted $\alpha$ oscillation component) and in Fig.~\ref{resolution}.

In Figs.~\ref{oscillations}b,c and~\ref{angles}, we extend this analysis to a second sample of oxygen concentration $x=$~0.54, for which oscillations are measured at several different angles $\theta$. Fits to Eqn.~\ref{fittwo} similar to those in Fig.~\ref{oscillations}a are made to the data in Fig.~\ref{angles}b as a function of angle (here $\Delta F_{\alpha,0}\lesssim~10$~T, $\Delta F_{\gamma,0}=74\pm 4$T, $F_{\alpha,0}\approx~535\pm 2$~T and $F_{\gamma,0}\approx~533\pm 2$~T). Fig.~\ref{angles}c shows the residual $\gamma$ oscillations on subtracting the fitted $\alpha$ component (green lines in Fig.~\ref{angles}b), and Fig.~\ref{angles}a shows the extracted peaks in the Fourier transform (shown in Fig.~\ref{oscillations}a and b and Appendix D) of the subtracted data (hollow symbols) and unsubtracted data (solid symbols). The angular dependent amplitude nodes observed in the residual $\gamma$ oscillations in Fig.~\ref{angles}c and frequencies in Fig.~\ref{angles}a are compared with those expected for a single corrugated cylinder, by using Yamaji's expression for an angular dependent frequency difference between the `neck' and `belly' of a single corrugated cylinder~\cite{yamaji1}:
\begin{equation}\label{yamaji}
\Delta F_{i,\theta}=\Delta F_{i,0}{\rm J}_0(k_\|c\tan\theta).
\end{equation}
On substituting the fit value of $\Delta F_{\gamma,0}$ into Eqn.~\ref{yamaji}, the predicted nodes for a single corrugated cylinder in Eqn.~\ref{nodeequation} are shown by lines in Fig.~\ref{angles}c, and predicted neck and belly frequencies are shown by red lines in Fig.~\ref{angles}a. We see that excellent agreement with the data is obtained in both Figs.~\ref{angles}a and c independent of any fitting parameters, confirming a single corrugated $\gamma$ Fermi surface section (distinct from $\alpha$). 

The small value of $\Delta F_{\alpha}$ implies that the putative nodes for the $\alpha$ Fermi surface section occur at fields too low for interference between neck and belly frequencies to be observed in the form of nodes. We can now understand why the $\alpha$ cylinder yields a spectrally dominant peak at $F_{\alpha,0}\approx$~532~T in Fig.~\ref{oscillations}a$-$ no significant smearing of the quantum oscillation phase occurs on averaging over $k_z$ for this section (see diagram in Fig.~\ref{pockets}a).
The larger corrugation of the $\gamma$ section, by contrast, implies that significant phase smearing occurs on averaging over $k_z$ for this section, with phase coherence being achieved only at the extrema $F_{\gamma,{\rm neck}}$ and $F_{\gamma,{\rm belly}}$$-$ accounting for the spectrally weak satellite frequencies flanking the central peak in Fig.~\ref{oscillations}a.

For a final consistency check, we turn to the harmonics  $2F_\alpha$, $2F_{\gamma,{\rm neck}}$ and $2F_{\gamma,{\rm belly}}$, which are observed for both samples ($x=$~0.54 and $x=$~0.56) in Figs.\ref{oscillations}a, b and c (plotted versus $\theta$ in Fig.~\ref{angles}a together with the Yamaji prediction). Because the harmonics appear in the Fourier transforms of the raw data (Figs.~\ref{oscillations}a, b and c and Appendix D), they provide a second estimate of each frequency (Fig.~\ref{oscillations}a) that is independent of any subtraction of the predominant $\alpha$ frequency and therefore independent of the previous fits. A further advantage of harmonic detection is that the doubling of the frequency yields a twofold increase in the confidence to which the separate Fermi surface sections can be resolved. The absolute frequency resolution obtained from the interval in magnetic field is therefore $\Delta F_{\rm lim}\approx[1/\Delta(1/\mu_0H)]/2\approx$~26~T. The lack of a peak splitting or deviation of
$F_\alpha$ and its harmonic from $F_\alpha=F_{\alpha}/\cos\theta$ over the entire angular range in Fig.~\ref{angles}a implies an irresolvably small depth of corrugation $\Delta F_{\alpha,0}\lesssim$~13~T, while the values of $\Delta F_{\gamma,\theta}=(2F_{\gamma,{\rm belly}}-2F_{\gamma,{\rm neck}})/4$ coincide well with the expected values from the fundamental analysis (shown in Fig.~\ref{angles}a). The observation of prominent harmonics for both the $\alpha$ and $\gamma$ pockets also enables us to estimate the relative size of the damping term for each of the electron and hole pockets, which we find to be similar in magnitude: $\Gamma_{\alpha}~\sim~\Gamma_{\gamma}~\sim~10^2$~T, corresponding to a mean free path of $\sim$~200~\AA.

We therefore find the nodal analysis of the oscillations, Fourier transform analysis, Yamaji fits and second harmonic detection {\it all} to yield internally consistent results corresponding to two similarly-sized $\alpha$ and $\gamma$ sections with significantly different degrees of corrugation. Oscillations from two pockets with the same degree of corrugation, as expected for bilayer split pocket areas in the underdoped cuprates~\cite{audouard1}, on the other hand, would not be consistent with our data (see Appendix G).

\subsection{Correspondence with equal and opposite electron and hole pockets}
\begin{figure}[ht!]
\begin{center}
\includegraphics*[width=.85\textwidth]{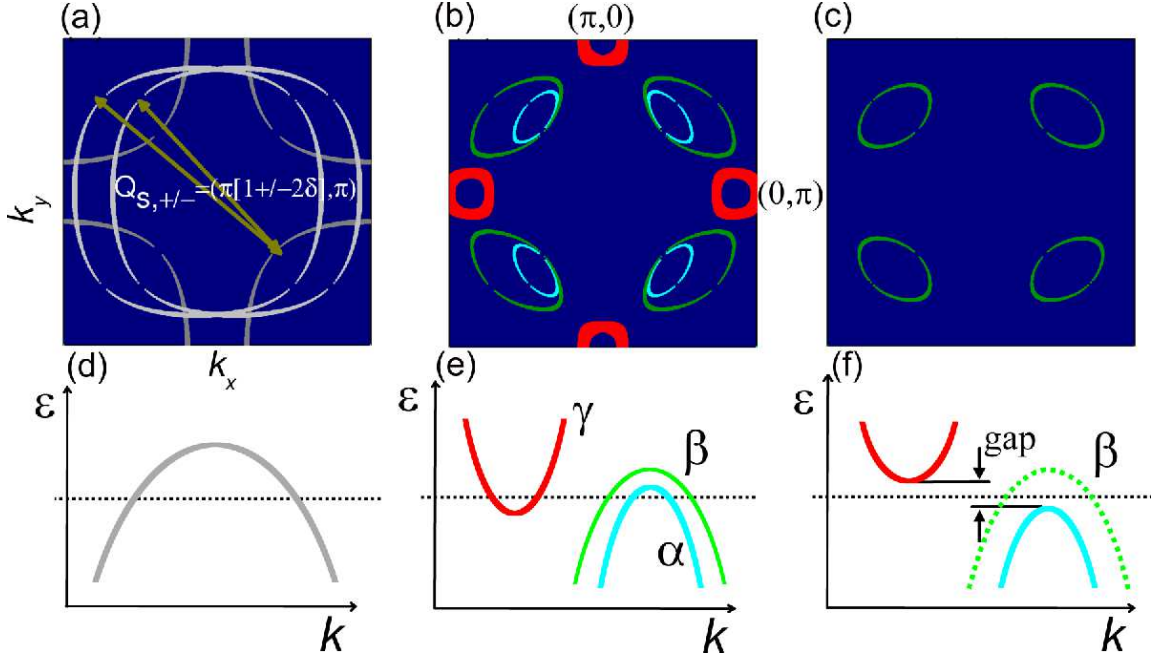}
\caption{{\bf a}, Schematic unreconstructed Fermi surface depicting the translational vectors ${\bf Q}_{\rm s,\pm}$, the line thickness (obtained by a perturbation in $r$ as described in the text) represents the inter-layer dispersion $k_z$ responsible for corrugation and is seen to be greater in the antinodal ($\pm\pi$,0), (0,$\pm\pi$) compared to the nodal ($\pm\pi/2$,$\pm\pi/2$) region. {\bf b}, Representative reconstructed Fermi surface according to models used in Refs.~\cite{dimov1,podolsky1,sebastian2,newkee} (described in the text and Appendices F, I). Blue and red lines correspond to the nodal hole pockets and antinodal electron pocket respectively. The line thickness (obtained by a perturbation in $r$) represents the inter-layer dispersion $k_z$ and is significantly bigger for the antinodal compared to the nodal pocket. {\bf c}, Reconstructed Fermi surface consisting of only a single type of pocket ($\beta$) after increasing the coupling $\Delta_{\rm s}$ (neglecting $\Delta_{\rm c}$). Antiferromagnetic bilayer coupling~\cite{andersen1} (shown in Fig.~\ref{bilayerfig} of Appendix F) would yield two slightly different variants of the Fermi surfaces in {\bf b} and {\bf c}, but with very similar pocket areas.
{\bf d}, Schematic dispersion prior to Fermi surface reconstruction depicted in one-dimension. {\bf e}, Schematic dispersion of the three pockets.
{\bf f}, Schematic showing the opening of a gap between the $\alpha$ and $\gamma$ pockets, as expected to result from an excitonic insulator instability. Here, we neglect the charge potential $\Delta_{\rm c}$, the effect of which on the $\beta$ orbit (or remnants thereof, shown as a dotted line) will depend on its strength and on the specifics of the superlattice (see Appendix J). The effects of bilayer splitting are not shown here for clarity, but are discussed in Appendix F.
} \label{OKfig}
\end{center}
\end{figure}

In the simplest model, our finding of Fermi surface sections with significantly different degrees of corrugation locates them at different locations in the Brillouin zone. We compare the different measured interlayer hoppings with calculated values in a tight binding model of underdoped YBa$_2$Cu$_3$O$_{6+x}$~\cite{xiang1} to locate the different pockets. As shown in Refs.~\cite{andersen1,pavarini1} the low-energy in-layer dispersion $\left[
\mathbf{k\equiv}\left(  k_{x},k_{y}\right)  \right]  $ of the Cu
$d_{x^{2}-y^{2}}-$like band is:%
\begin{equation}
\varepsilon\left(  \mathbf{k}\right)  =-4t\left[  u\left(  \mathbf{k}\right)
+2r\frac{v\left(  \mathbf{k}\right)  ^{2}}{1-2ru\left(  \mathbf{k}\right)
}\right]  
\label{OK1}
\end{equation}
where $u(\mathbf{k})\equiv\frac{1}{2}(\cos k_{x}+\cos k_{y})$,  $v(\mathbf{k})\equiv\frac{1}{2}(\cos k_{x}-\cos k_{y})$, $t$ is the nearest neighbour hopping integral, and $r$ is a `range' parameter due to in-layer hopping via the so-called axial
orbital~-~a hybrid of mostly Cu 4$s$ and apical oxygen 2$p_{z}$. For small $r,$ Eqn.~\ref{OK1} reduces to the familiar tight-binding form:%
\begin{align}
\varepsilon\left(  \mathbf{k}\right)    &=const.-2t\left(  \cos k_{x}+\cos
k_{y}\right)  +4t^{\prime}\cos k_{x}\cos k_{y}\nonumber\\
&-2t^{\prime\prime}\left(  \cos2k_{x}+\cos2k_{y}\right)  
\label{OK2}
\end{align}
with $t^{\prime}=rt$,  and $t^{\prime\prime}=\frac{1}{2}rt$. Hopping in the $z$-direction occurs via the axial orbital and therefore
depends on $\left(  k_{x},k_{y}\right)  $ as $\frac{v\left(  \mathbf{k}%
\right)  ^{2}}{1-2ru\left(  \mathbf{k}\right)  },$ (i.e.) it has the same symmetry as the superconducting gap, vanishing along the nodal lines and reaching maxima
at the $\left(  \pm\pi,0\right)  $ and $\left(  0,\pm\pi\right)  $ points. If
the layers are stacked on top of each other, the $k_{z}$-dispersion is simply
included in Eqn.~\ref{OK1} by the substitution: $r\rightarrow r+\left(  t_{\perp
}/t\right)  \cos ck_{z}.$ For YBa$_2$Cu$_3$O$_{6+x}$ ($x \approx 0.5$), $t \approx 400$~meV, $t_\perp$ causes $r$ to vary between $\approx$ 0.28 and 0.32 and the corresponding Fermi surface is shown in Fig.~\ref{OKfig}a. Here the line thickness is the size of the $k_z$ dispersion yielded by the perturbation in $r$. The different interlayer hoppings at different locations in the Brillouin zone reveals that small pockets at each of these locations created by Fermi surface reconstruction will have different corrugations. Accordingly, we proceed to examine the difference in corrugation expected for each of the multiple pockets throughout the Brillouin zone yielded for a translational symmetry breaking order parameter. 

For illustrative purposes, we adopt a model involving a single transformation of $\varepsilon_{\bf k}$ by ${\bf Q}_{\rm s,\pm}=(\pi[1\pm2\delta],\pi)$ of the helical spin-density or d-density wave type such as that used in Refs.~\cite{dimov1,podolsky1,sebastian2,newkee} as one possibility, with an incommensurability $\delta\approx\frac{1}{16}\approx$~0.06 to match incipient spin order and / or excitations~\cite{haug1,stock1,stock2}), and a single variable $\Delta_{\rm s}$ adjusted to have the value $0.625 t$, as explained in the next section. The Fermi surface thus obtained comprises equal-sized electron (red) and hole (blue) pockets at the antinodal and nodal positions respectively, accompanied by a larger (green) hole pocket at the node (shown in Fig.~\ref{OKfig}b). Here again, the line thickness is the size of the interlayer hopping. A significant difference in corrugation (resulting from the dispersion in $k_z$) is expected for pockets located at the nodal and antinodal locations: the obtained interlayer corrugation for each of the smaller and larger nodal pockets and the antinodal pocket is of order 14, 45, and 180 T respectively, being representative of general translational symmetry breaking order parameters yielding pockets at the antinodal and nodal regions of the Brillouin zone~\cite{dimov1,podolsky1,sebastian2,newkee, millis1,harrisonSDW}. The difference in corrugation thus provides an opportunity for their respective locations to be identified in quantum oscillation measurements. We find that the ratio of observed corrugations for the $\alpha$ and $\gamma$ pocket is comparable to that expected for the nodal and antinodal pockets. The association of the nodes and antinodes with the expected positions of hole and electron pockets within a broken symmetry Fermi surface~\cite{dimov1,podolsky1, newkee}, further identify $\alpha$ as a two-dimensional hole pocket and $\gamma$ as a corrugated electron pocket within this picture. Additionally, bilayer splitting in the case where bilayers are ferromagnetically coupled is expected to result in a large splitting of frequencies, as opposed to the case of antiferromagnetic coupling between bilayers, in which case the frequency splitting of the bilayers would remain small. The lack of an observed splitting in any of the $\alpha$, $\beta$, and $\gamma$ frequencies is therefore indicative of antiferromagnetic coupling between bilayers (see Appendix F for a discussion of bilayer splitting). The observation in inelastic neutron scattering experiments of low energy spin excitations and a spin resonance near $(\pi,\pi)$ that antiferromagnetically couple bilayers~\cite{fong1} suggests the association of spin order with pocket formation.

For a larger $\beta$ orbit corresponding to a distinct section of Fermi surface (as opposed to a magnetic breakdown orbit)~\cite{sebastian2}, its character can be identified in a scenario where $\alpha$ and $\gamma$ correspond to equal and opposite hole and electron pockets at nodal and antinodal locations. Luttinger's theorem~\cite{luttinger1} (requiring a hole concentration of $p=$~10~$\pm$~1~\%~\cite{liang1}) and the absence of a discernible splitting of $F_\beta$ from corrugation, suggests the association of the $\beta$ orbit with a larger `hole-like' pocket located near $(\pi/2,\pi/2)$ (see Fig.~\ref{pockets}a). The entire hole doping concentration would then be provided by the $\beta$ pocket, which acts as a reservoir of charge carriers, while the $\alpha$ and $\gamma$ pockets almost cancel out in carrier concentration.

\subsection{Electronic origin of the observed pockets}
In the simplest case, our observations indicate a broken translational symmetry order parameter that creates differently corrugated Fermi surface pockets at different locations in the Brillouin zone - Fermi surface models that yield pockets only at the nodal locations would face a considerable challenge in explaining our observations.

Spin ordering involving a broken translational symmetry has been reported in YBa$_2$Cu$_3$O$_{6+x}$ samples with oxygen concentrations $x\approx$~0.45~\cite{haug1,keimer1} and $\approx$~1.0~\cite{mitrovic1} in an applied magnetic field. Antiferromagnetic order has not yet been observed in samples with oxygen concentrations (0.49~$<x\lesssim$~0.8) in which quantum oscillations are observed~\cite{doiron1,audouard1,sebastian2,sebastian1,singleton1} except when 2$\%$ of the Cu ions are replaced by spinless Zn ions in samples with $x~\approx~0.6$~\cite{keimer2}. The strong suppression of Zeeman splitting of the Landau levels has also been reported in ref.~\cite{sebastian3} with additional evidence (Appendix C) provided from the doping dependent~\cite{sebastian1} and harmonic oscillations resolved in this work, suggesting a translational symmetry breaking order parameter involving spin degrees of freedom or alternate means of suppression of the spin degrees of freedom. In addition, the implicit near-degeneracy of the bilayer split pocket frequencies (see Fig.~\ref{angles}a and Appendices F and G) signal antiferromagnetically coupled bilayers.

\begin{figure}[ht!]
\begin{center}
\centerline{\includegraphics*[width=.75\textwidth]{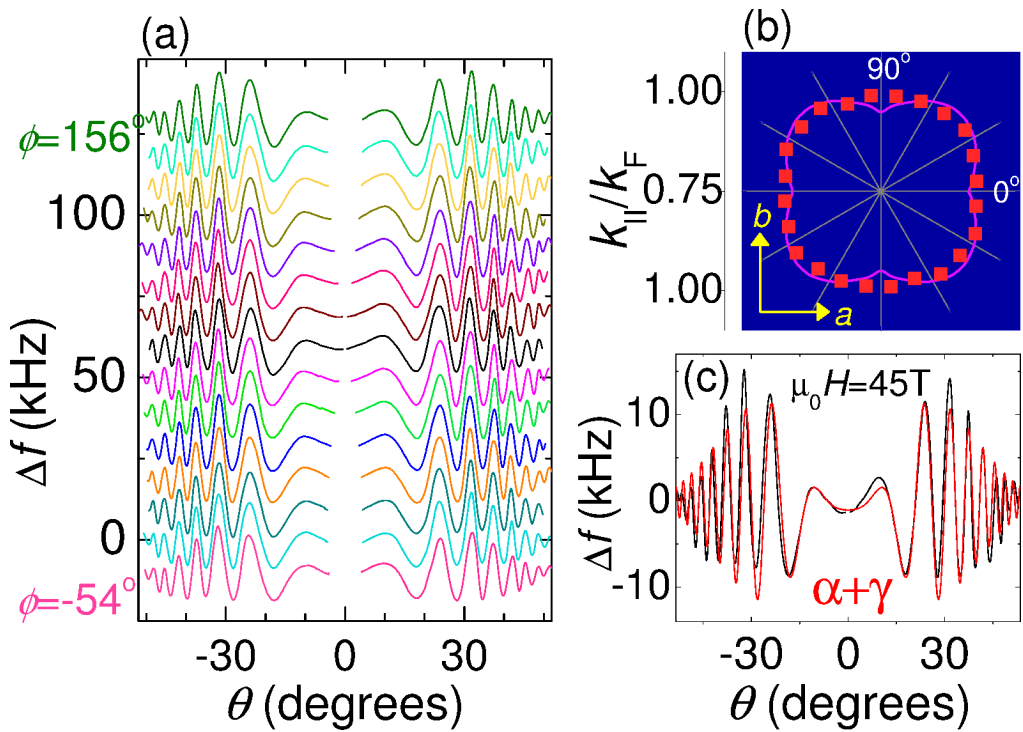}} 
\caption{{\bf a}, Examples of angle-swept measurements made from positive to negative values of $\theta$ on a sample of
YBa$_2$Cu$_3$O$_{6+x}$ with $x=0.56$ at fixed field $\mu_0H=$~45~T, for different in-plane orientations ($\phi$) of the in-plane component of the
magnetic field ${\bf H}_\|=(H\sin\theta\cos\phi,H\sin\theta\sin\phi,0)$. Curves shown correspond to $-$54~$\leq\phi\leq$~156$^\circ$ (bottom to
top) in steps of 15$^\circ$. The gap in angular data around $\theta=0$ is due to a correction made for a small misalignment of the crystal, explained in Appendix A. {\bf b}, The $\phi$-dependence of the caliper radius $k_\|$ of the $\gamma$ pocket determined from fits (see Appendix E) of superimposed $\alpha$ and $\gamma$ oscillation waveforms (e.g. Eqn.~\ref{fittwo})
 to the data in ({\bf a})$-$ with $\Delta F_{\gamma,0}$ and $\Gamma_\alpha$ held constant at values obtained
from fits to the field swept data. Fit frequencies are given in Appendix E. The full 360$^\circ$ $\phi$ rotation is inferred from
performing a rotation over 180$^{\circ}$ in $\phi$ combined with fits to $\pm\theta$, with only $k_\|$ allowed to vary. The magenta line is the $\phi$-dependence of $k_\|$ expected for the single ${\bf Q}_{\rm s,\pm}$ model shown in Fig.~\ref{OKfig}b. {\bf c} An example fit (red
line) to $\phi=$~51$^\circ$ (equivalent to 231$^\circ$) data (black line), with further fits shown in Appendix E.}
\label{twoaxis}
\end{center}
\end{figure}

To better understand the underlying origin of the electronic structure, we map the in-plane topology of the corrugated pocket. The in-plane topology of the warped $\gamma$ pocket is reflected in a variation of the in-plane calliper radius $k_\|$ (mapped, for example in Ref.~\cite{nam1}, see Appendix K). In our experiment, the calliper radius is mapped by two-axes angle-dependent quantum oscillation measurements where $\theta$ (the angle between ${\bf H}$ and the $c$-axis) is swept while $H=|{\bf H}|$ remains fixed, the sweeps being repeated for many different orientations $\phi$ of the in-plane component of the magnetic field ${\bf H}_\|=(H\sin\theta\cos\phi,H\sin\theta\sin\phi,0)$.  Because of its low degree of corrugation, the method described here is not applicable for determination of the in-plane Fermi surface topology of the $\alpha$ hole pocket. The significant corrugation of the $\gamma$ pocket, by contrast, enables variations in $k_\|$ to be mapped in Fig.~\ref{twoaxis}. The quantum oscillation data as a function of $\theta$ at each value of $\phi$ is fit to a waveform composed of superimposed $\alpha$ and $\gamma$ oscillations as described in Appendix E, and the calliper radius thereby extracted at various values of $\phi$. Our results show a rounded-square in-plane topology of the $\gamma$ pocket. 

The rounded-square (i.e. nearly circular) topology of the $\gamma$-pocket (shown in Fig.~\ref{twoaxis}b) is broadly consistent with an electron pocket predicted by numerous Fermi surface reconstruction models~\cite{dimov1,podolsky1,sebastian2,newkee,millis1}, although unconventional models - for example those involving novel quasiparticles - cannot be ruled out. The coexistence of a near circular electron pocket with hole pockets of comparable and/or larger size does not appear consistent with presently proposed `stripe' models involving multiple translations of the electronic bands $\varepsilon_{\bf k}$ by ordering vectors of the form ${\bf Q}=(\pi[1\pm2\delta],\pi)$ where 0~$<\delta\lesssim\frac{1}{8}$~\cite{millis1,harrisonSDW,hackl1}. We consider an example of translational symmetry breaking model that could yield pockets of the observed size and shape to facilitate an estimation of the size of the order parameter $\Delta_{\rm s}$ that breaks translational symmetry. The model we consider involves a single transformation of $\varepsilon_{\bf k}$ by ${\bf Q}_{\rm s,\pm}=(\pi[1\pm2\delta],\pi)$ of the helical type described in the earlier section. 
Adjusting the value of a single parameter $\Delta_{\rm s}=0.625 t$ in such a model yields three pockets of size (to within 2$\%$), carrier type, corrugation depth and shape consistent with experiment$-$ see Fig.~\ref{OKfig} and Appendix I, in conjunction with the
experimentally observed constraint of hole doping $p\approx$~11.7~\% (see Fig.~\ref{OKfig}) and $\delta\approx\frac{1}{16}$). The large ratio of $\Delta_{\rm s}/t$ indicates a
substantial coupling, consistent with the near-circular topology of the electron pocket (the small magnitude of deviation from a circle is seen
from Fig.~\ref{twoaxis}b). While band structure calculations indicate $t\approx$~400~meV~\cite{andersen1}, $t\approx$~100~meV provides closer correspondence with the observed effective masses (see Fig.~\ref{oscillations}, lower inset), suggesting $\Delta_{\rm s}\sim$~60~meV. Such a large value of $\Delta_{\rm s}$ is challenging to reconcile with the absence of signatures of long range order in neutron scattering experiments in YBa$_2$Cu$_3$O$_{6+x}$ samples of composition 0.49~$<x\lesssim$~0.8 (where quantum oscillations are observed). An unconventional form of broken translational symmetry-breaking spin order are suggested, including staggered moments present chiefly in the vortex cores~\cite{mitrovic1}, magnetic field-stabilised spin density wave order~\cite{keimer1}, or spin ordering with d-wave pairing symmetry~\cite{dimov1,podolsky1}; although unconventional models not involving translational symmetry breaking cannot be ruled out.

\section{III. Discussion}

An open question pertains to the origin of the unexpected similarity in size of the $\alpha$ and $\gamma$ pockets observed in underdoped YBa$_2$Cu$_3$O$_{6+x}$. More information on the microscopic nature of the primary instability will assist in addressing this question. Irrespective of their origin, the existence of compensated pockets suggests a Fermi surface prone to a secondary instability~\cite{kohn1,halperin1} at which the compensated pockets are destroyed. 

\subsection{Excitonic instability}
The possibility of an `exciton insulator' instability (at which the compensated $\gamma$ (electron) and $\alpha$ (hole) pockets are destroyed) is raised by their similar volumes (within the experimental uncertainty, both occupying 1.91~$\pm$~0.01~\% of the Brillouin zone) with effective masses sufficiently large ($m^\ast_\alpha\approx m^\ast_\gamma\approx$~1.6~$m_{\rm e}$ for $x=$~0.56, found from the fit in the lower inset to Fig.~\ref{oscillations}a) to cause strong attraction. 
Strong coupling between pairs of electrons and holes is suggested by a binding energy ($Ryd^\ast\sim$~430~meV, derived in the Appendix B) that
is more than ten times larger than their individual kinetic energies ($E_{\rm ke}\sim$~39~meV). This also means that the putative exciton separation to Bohr radius ratio is larger than unity (i.e. $r_{\rm s}\approx$~3.4 for $x=0.54$, see Fig.~\ref{pockets}b), rendering the system unstable to an excitonic
insulating instability~\cite{littlewood1}. Due to the strongly interacting dilute Bose-Einstein condensate (BEC) limit in which this system is consequently positioned (i.e. $r_{\rm s}\gg$~1), the excitonic attraction is not contingent on exactly matching pocket shapes or perfect `nesting.'

Suggestive of an instability actually taking place is the occurrence of a metal-insulator quantum critical point (QCP) at $x_{\rm c}=6.46$ identified in recent experiments~\cite{sebastian1,ando1}. The threefold increase in effective mass on reducing $x=$~0.54 to $x=$~0.49 immediately preceding the QCP signals a tenfold increase in $Ryd/E_{\rm ke}$ to $\sim$~10$^2$ (see Appendix B) and threefold increase in $r_{\rm s}$ to $\sim$~10 on approaching the critical doping (shown in Fig.~\ref{pockets}b), potentially indicating a further reduction in the effectiveness of screening of the Coulomb attraction, with the final trigger for an excitonic instability occurring at $x_{\rm c}$. Strongly correlated systems that support conditions propitious for an excitonic insulator are unusual~\cite{kohn1,jerome1,littlewood1}, given that a single carrier type rather than equal and opposite types typically results close to a Mott insulating state~\cite{fazekas1}. This situation appears to be reversed in YBa$_2$Cu$_3$O$_{6+x}$, where Fermi surface reconstruction at the first instability transforms the single band of carriers with a strong on-site repulsion into two bands containing equal and opposite pockets that are strongly attracted to each other (illustrated in Fig.~\ref{OKfig}a,b). 

\subsection{Destruction of compensated pockets}

A secondary superlattice instability would be anticipated to accompany exciton condensation for $x<x_{\rm c}$~\cite{halperin1}, destroying the compensated $\alpha$ and $\gamma$ pockets (see Figs.~\ref{OKfig}d and e). Drawing an analogy with other underdoped cuprates, accompanying charge order could be expected at wavevectors ${\bf Q}_{\rm c}=2{\bf Q}_{\rm s}$~\cite{kivelson1} and / or ${\bf Q}_{\perp}=(0,\pi/2)$~\cite{shen1} (see Appendix J).
The possibility
of the charge plus spin character of the second instability is suggested by the wavevector match ${\bf Q}_{\rm c}=2{\bf Q}_{\rm s}$ between the
`charge' wavevector ${\bf Q}_{\rm c}$ characterizing phonon broadening~\cite{mook2} and possible superlattice formation~\cite{mook1}, and the
`spin' wavevector ${\bf Q}_{\rm s}$ associated with short range spin order~\cite{haug1} and magnetic excitations~\cite{stock1}. A finite charge superlattice potential $\Delta_{\rm c}$ would reconstruct the remaining $\beta$ pocket into smaller pockets or open sheets~\cite{millis1}$-$ the precise details depending on the choice of wavevector (see Appendix J).
 
Further, the emergence of signatures of quasi-static magnetism in neutron scattering~\cite{keimer1,haug1} and and local moment behavior in muon spin rotation experiments~\cite{sanna1,sonier1} is consistent with a strengthened $\Delta_{\rm s}$ for the region $x<x_{\rm c}$. A strengthened $\Delta_{\rm s}$ alone could gap the $\alpha$ and $\gamma$ pockets in YBa$_2$Cu$_3$O$_{6+x}$ (see Fig.~\ref{OKfig}c), leaving behind only the $\beta$ hole pockets at the nodes~-~reminiscent of strong coupling models~\cite{shraiman1}. 

\subsection{Reconciliation with photoemission experiments}
Signatures from our measurements that the $\alpha$ pocket from which predominant oscillations arise is of the `hole' carrier type located at the nodes, brings the results of quantum oscillation measurements close to reconciliation with the results of angle-resolved photoemission spectroscopy (ARPES) experiments. A dichotomy arises at the antinodes, however; while quantum oscillations indicate antinodal Fermi surface pockets measured by quantum oscillations, a large antinodal `pseudogap' has been measured by ARPES measurements. An intriguing question concerns whether the finite spectral weight measured by ARPES to be concentrated at `arcs' of finite extent centred at the nodal points $(\pm\pi/2,\pm\pi/2)$ could conceivably reflect the sole remaining parts of the $\beta$ Fermi surface on annihilation of the equal and opposite $\alpha$ and $\gamma$ hole and electron pockets at the secondary Fermi surface instability as shown in Fig.~\ref{OKfig}~\cite{hossain1} $-$ a Fermi surface consisting of `arcs' is reported in materials with robust stripe order~\cite{chang1}. While such an interpretation is suggestive, an additional explanation is still required for the closing of the ARPES pseudogap in the low temperature and high magnetic field regime relevant to quantum oscillation measurements.

One possibility suggested by recent models~\cite{senthil1,micklitz1} for the re-emergence of spectral weight at the antinodes in quantum oscillation measurements is that incoherent superconducting pairs destroy the spectral weight under the experimental conditions ($T\gtrsim T_{\rm c}$ and $B=$~0) required for photoemission experiments, while this spectral weight is recovered in the `quantum vortex liquid' regime accessed by quantum oscillation measurements. Another contribution could arise from fluctuations of the excitonic insulating instability that cause loss of antinodal spectral weight over a range of oxygen concentrations above $x_{\rm c}$ extending to higher temperatures$-$ similar to the case of NbSe$_2$ and TiSe$_2$ above their charge ordering temperatures~\cite{borisenko1,monney1}. The possible role of density wave reconstruction has also been suggested by theory~\cite{chubukov1} and by results of ARPES experiments~\cite{hashimoto1, campuzano1}. Alternatively, more exotic models involving novel quasiparticles visible to quantum oscillations and not ARPES have also been proposed~\cite{gailitski1}.

\subsection{Relevance to pairing}
Finally, we consider the effect that excitations of the potential excitonic insulator instability may have on the quasiparticle pairing mechanism. The unusual occurrence of a Fermi surface ideally predisposed to an excitonic insulator instability in the same material where high $T_{\rm c}$ superconductivity occurs could either suggest an link between the phenomena, or a coincidence. Similar compensated corrugated cylindrical Fermi surface pockets are found in other families of unconventional superconductor that also exhibit unexpectedly high transition temperatures$-$ such as the Fe pnictides~\cite{dong1} and Pu-based superconductors~\cite{plut}.

Theoretical proposals made prior to the advent of high temperature superconductivity considered conditions under which an excitonic mechanism may
potentially enhance $T_{\rm c}$~\cite{allender1}. Whether or not such a cooperative mechanism is in play in the d-wave cuprate superconductors,
however, is contingent on a better understanding of the relative importance of spin and charge fluctuations in contributing to high $T_{\rm c}$
in these materials.

\section{Appendix}

\subsection{A. Experimental details}

Quantum oscillations are measured using the contactless conductivity method~\cite{coffey1} in two pulsed magnets reaching 65~T and 85~T, and in a DC magnet reaching 45~T. Plate-like single crystals of YBa$_2$Cu$_3$O$_{6+{\rm x}}$~\cite{liang1} with in-plane dimensions 0.3~$\times$~0.8~mm$^2$ are attached to the face of a flat 5 turn coil (compensated in the case of pulsed field experiments) that forms part of a proximity detector circuit resonating at $\sim$~22~MHz~\cite{altarawneh1}. A change in the sample skin depth (or complex penetration depth in the superconducting state~\cite{coffey2}) leads to a change in the inductance of the coil, which in turn alters the resonance frequency of the circuit. Quantum oscillations are observed in the in-plane resistivity and hence the skin depth).

To minimize the effects of flux dissipation heating in pulsed magnetic fields, the sample is immersed in liquid $^4$He throughout the experiment~\cite{fritz1,corcoran1}, and only the oscillations observed during the falling field are considered for detailed analysis. On taking such precautions, effective mass estimates are obtained that are comparable to those obtained in samples of the same oxygen concentration in static magnetic fields~\cite{sebastian2}.

The fixed $H$ ($=$~45~T) angle swept $\theta$ measurements performed at different in-plane angles $\phi$ shown in Fig.~\ref{twoaxis} are obtained using a dual-axis rotator. Due to a small misalignment of the sample by $\theta_0\approx$~7$^\circ$ at $\phi_0\approx$~51$^\circ$, a minor correction $\cos\theta_{\rm actual}=\cos\theta_{\rm uncorrected}\times\sqrt{\cos^2\theta_0+\sin^2\theta_0\sin^2(\phi-\phi_0)}$ is made to obtain the correct $\theta$.

\subsection{B. Estimates of the exciton binding energy}

We estimate the effective `Rydberg' for exciton binding to be~\cite{jerome1,littlewood1} 
\begin{equation}\label{binding}
Ryd^\ast=\frac{\mu}{m_{\rm e}}\frac{1}{\epsilon^2}Ryd 
\end{equation}
$\approx$~430~meV for YBa$_2$Cu$_3$O$_{6+x}$ samples of oxygen concentration $x=$~0.54 and 0.56~\cite{liang1}, where $\mu\approx m^\ast/2\approx$~0.8~$m_{\rm e}$ is the reduced bi-exciton mass, $Ryd\approx$~13.6~eV is the hydrogen Rydberg and $\epsilon\approx$~5 is the relative permittivity for YBa$_2$Cu$_3$O$_{6+x}$~\cite{humlicek1}. The kinetic energy is estimated using 
\begin{equation}\label{kineticenergy}
E_{\rm ke}=\frac{e\hbar F}{m^\ast}
\end{equation}
$\approx$~39~meV for $F=535$T and $m^{\ast}=$~1.6~$m_{\rm e}$, yielding a ratio $Ryd^\ast/E_{\rm ke}\approx$~11. The effective Bohr radius of the exciton is estimated to be 
\begin{equation}\label{bohrradius}
a^\ast=\epsilon\frac{m_{\rm e}}{\mu}a_0
\end{equation}
$\approx$~3.3~\AA, where $a_0\approx$~0.53~\AA~ is the hydrogen Bohr radius. We estimate the exciton density as $n=A_k/2\pi^2\approx$~2.6~$\times$~10$^{17}$~m$^{2}$, where $A_k=2\pi eF/\hbar\approx$~5.1~$\times$~10$^{17}$~m$^{-1}$, yielding an effective exciton separation of
\begin{equation}\label{excitonseparation}
r_{\rm s}\approx\frac{1}{a^\ast}\sqrt{\frac{1}{\pi n}}
\end{equation}
from which $r_{\rm s}\approx$~3.4.

In estimating $Ryd^\ast$ and $r_{\rm s}$, we have neglected the effects of screening, which could become significant should a large $\beta$ pocket survive the formation of a superstructure accompanying exciton condensation. On the other hand, superstructure formation (see following section) would likely cause significant reconstruction of the $\beta$ orbit$-$ the outcome being strongly dependent on the associated ordering vectors and nature of the broken symmetry (see Appendix J). 
Screening will also become ineffective if $m^\ast$ globally diverges at the quantum critical point at $x_{\rm c}\approx$~0.46, at which point both $Ryd^\ast$ and $r_{\rm s}$ would diverge while $E_{\rm ke}$ collapses to zero.

When the quasiparticle effective mass changes (as reported in YBa$_2$Cu$_3$O$_{6+x}$ as a function of doping~\cite{sebastian1}), then we can
expect the following dependences (shown in Fig.~\ref{pockets} of the main text):
\begin{equation}\label{dependences}
\frac{Ryd^\ast}{E_{\rm ke}}\propto (m^{\ast})^2\hspace{0.5cm}{\rm and}\hspace{0.5cm}r_{\rm s}\propto m^\ast
\end{equation}
indicating that tuning of the electron correlations is expected to have dramatic consequences for the likelihood of exciton pairing and the
applicability of the strong coupling BEC limit~\cite{littlewood1,kohn1}.

\subsection{C. Extended analysis of spin splitting}
\begin{figure*}[h!]
\centering
\begin{center}
\includegraphics*[width=.9\textwidth]{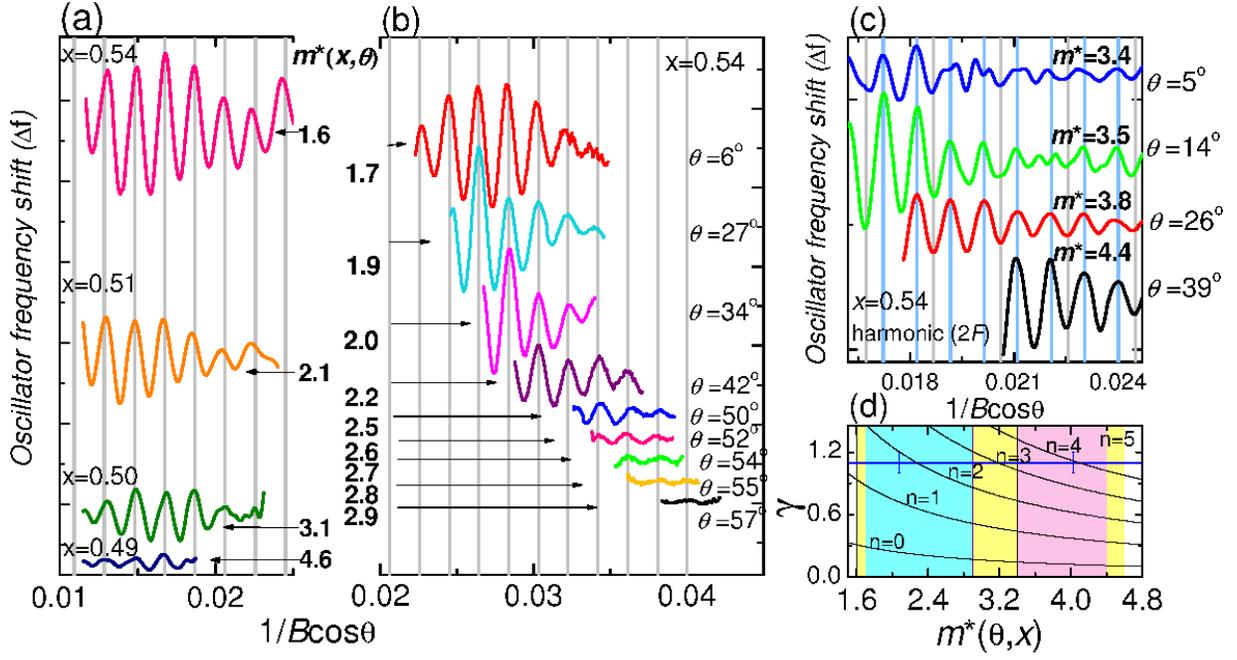}
\caption{{\bf a} Fundamental quantum oscillations measured for a range of effective masses tuned by doping, taken from ref.~\cite{sebastian1}. Here the quantum oscillation amplitude has been rescaled by an exponential factor for purposes of comparison. {\bf b} Fundamental quantum oscillations measured for different effective masses tuned by angular rotation, taken from ref.~\cite{sebastian3}. Each set of quantum oscillations is labelled by the tuning parameter (angle, doping) and the corresponding effective mass marked by arrows. The gray vertical lines label the expected peak positions for the fundamental $\alpha$ quantum oscillations in {\bf b} and {\bf c}. {\bf c} Harmonic quantum oscillations measured for a range of effective masses tuned by angular rotation, obtained by subtracting the fundamental alpha and gamma oscillations from the overall signal measured in the current work (Fig.~\ref{angles}b). Vertical cyan lines label the expected peak positions for the harmonic (2$F$) $\alpha$ quantum oscillations. {\bf a}, {\bf b}, and {\bf c} reveal an absence of phase inversion of the measured quantum oscillations over a broad range in effective mass. {\bf d} The effective mass locations of the spin zeros expected for different values of the spin doublet moment $\gamma$. The value of $\gamma$ (i.e. the Wilson ratio) in the paramagnetic phase is marked by the horizontal blue line. The yellow shading represents the range of effective mass tuned by the doping in {\bf a}, the cyan shading the range of effective mass tuned by the angle in {\bf b}, and the pink shading the range of effective mass covered by the angle-tuned harmonics in {\bf c}. Given the absence of phase inversion (which would locate a spin zero) in quantum oscillations over the measured range of effective mass, $\gamma \ll 1$ is indicated for the spectrally dominant $\alpha$ pocket.} \label{spinzeros}
\end{center}
\end{figure*}

For a finite value of $\gamma$ (which is the effective moment of the spin doublet (i.e.) the Wilson ratio in the paramagnetic phase), the spin damping factor $R_{\rm s}=\cos(\pi\gamma m^\ast/m_{\rm e})$ has an oscillatory form. In the case of underdoped YBa$_2$Cu$_3$O$_{6+x}$, however, it has been shown in ref.~\cite{sebastian3} and in greater detail here, that the value of $\gamma\ll$~1 for the $\alpha$ pocket corresponding to the spectrally dominant oscillations, leading to a value $R_{\rm s} \approx 1$. The effect of an oscillatory $R_{\rm s}$ would be for the quantum oscillation amplitude to vanish at `spin zero' values of $\gamma m^\ast / m_{\rm e} = \left(2{\rm n}+1\right)/2$, on either side of which the phase of the quantum oscillations changes by $\pi$, resulting in their inversion. We test for an oscillatory $R_{\rm s}$ by examining the phase of measured quantum oscillations for a range of $m^\ast$ tuned by angular rotation~\cite{sebastian3} and by changing the doping~\cite{sebastian1}. Figure~\ref{spinzeros}a,b shows the phase of the fundamental oscillations to be invariant as a function of $m^\ast$ from 1.7 to 3.0 $m_{\rm e}$ on angular rotation, and from $m^\ast$ from 1.6 to 4.5 $m_{\rm e}$ on varying the doping. Figure~\ref{spinzeros}c shows the phase of the harmonic (2$F$) oscillations to be invariant as a function of $m^\ast$ from 3.4 to 4.4 $m_{\rm e}$ on angular rotation. The measured quantum oscillations dominated by the $\alpha$ frequency are seen to be phase invariant for a wide range in $m^\ast$. From figure~\ref{spinzeros}d, a finite $\gamma$ would lead to multiple spin zeros over the measured range in $m^\ast$, the absence of which signals a value of $\gamma \ll 1$ for the dominant $\alpha$ frequency. Consequently, $R_{\rm s} \approx 1$ does not appear in Eqn.~\ref{fittwo}.

\subsection{D. Extended analysis of swept magnetic field data}

\begin{figure*}[htbp]
\centering
\begin{center}
\includegraphics*[width=.7\textwidth]{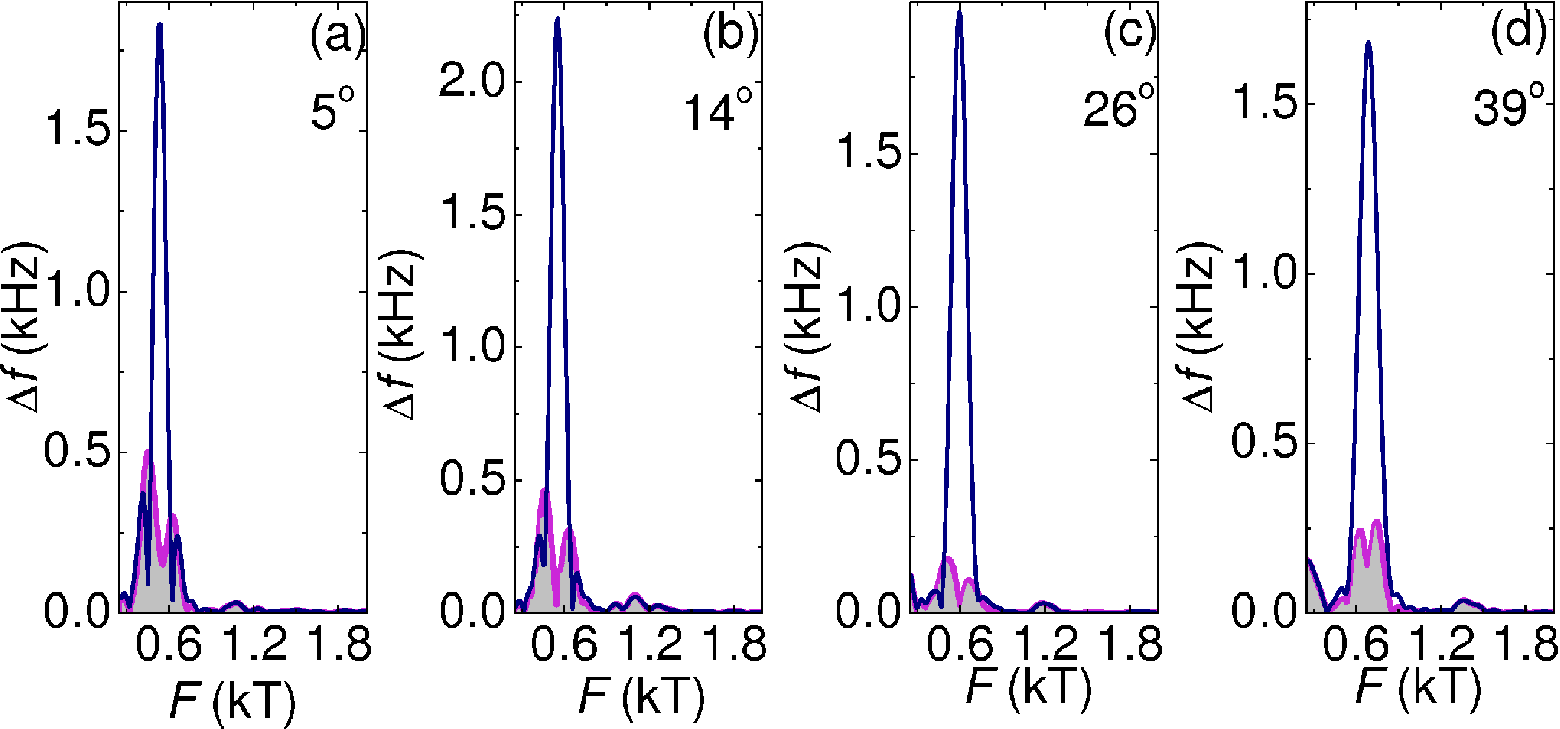}
\caption{{\bf a}, {\bf b}, {\bf c} and {\bf d}, Fourier transforms of the oscillations measured in YBa$_2$Cu$_3$O$_{6+x}$ with $x=$~0.54 at different angles $\theta$ before (mauve line) and after (magenta line with grey shading) subtracting the $\alpha$ component of a fit to Eqn.~\ref{fittwo} of the main text$-$ a subset of which are shown in Fig.~\ref{oscillations}b and c of the main text. The resolved $\gamma$ frequencies are used to construct Fig.~\ref{angles}a of the main text.} \label{ffts}
\end{center}
\end{figure*}

During Fourier analysis, weaker spectral features can be affected by interference effects between oscillations from a spectrally dominant series of oscillations (nearby in frequency) and the window function (a Hann window in the current study) used to modulate the data. Such interference between the $\alpha$ oscillation and the window function affects the resolved $\gamma$ frequencies in the present study. One way to mitigate interference is to subtract fits to a dominant well characterized sequence of oscillations (i.e. the $\alpha$ oscillations) prior to Fourier analysis. This procedure is justified in the present case owing to the resolved separation between the $\alpha$ and $\gamma$ frequencies in our Fourier analysis (see Fig.~\ref{angles}, Fig.~\ref{oscillations}, and Fig.~\ref{ffts})$-$ with the degree of separation growing proportionately larger for the harmonics as expected. 

In Fig.~\ref{ffts} we show Fourier transforms of the measured oscillations in samples with $x=$~0.54 before (i.e. Fig.~\ref{angles}b of the main text) and after (i.e. Fig.~\ref{angles}c of the main text) subtracting the fit of the $\alpha$ oscillations (Eqn.~\ref{fittwo} in the main text), a subset of which are shown in Figs.~\ref{oscillations}b and c of the main text. The $\gamma$ frequencies $F_{\gamma,{\rm neck}}$ and $F_{\gamma,{\rm belly}}$ in Fig.~\ref{angles}a of the main text are determined with greater accuracy by performing Fourier transformation after subtracting the $\alpha$ oscillations. 

\begin{figure*}[htbp]
\centering
\begin{center}
\includegraphics*[width=.7\textwidth]{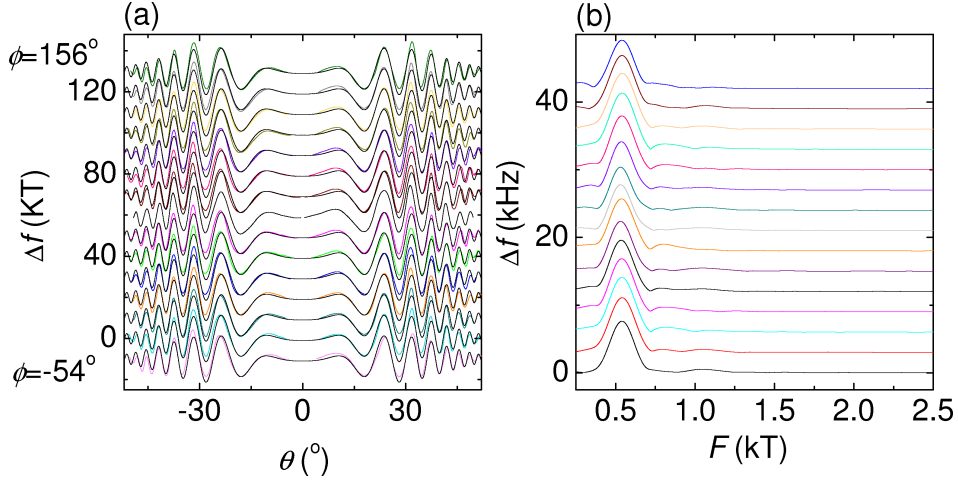}
\caption{{\bf a} Swept $\theta$ oscillations measured in a sample of YBa$_2$Cu$_3$O$_{6+x}$ with $x=$~0.56 at constant field $H=$~45~T and different fixed values of $\phi$ at $T\approx$~1.5~K fit to Eqn.~\ref{waveform} (black lines) as described in Fig.~\ref{twoaxis}. The gaps in the data result from a correction for a sample misalignment of $\approx$~7$^\circ$ (see experimental details). {\bf b} Fourier transforms in $1/\mu_0H\cos\theta$ of the oscillations shown in {\bf a}, which are dominated by the $\alpha$ pocket. As expected for a nearly ideal two-dimensional Fermi surface section, the dominant $\alpha$ frequency obtained is the same as that from fixed $\theta$ swept magnetic field experiments (i.e. Fig.~\ref{twoaxis}a) at all angles of rotation. The curves correspond from bottom to top to $-$54~$\leq\phi\leq$~156$^\circ$ in 15$^{\circ}$ steps.} \label{fits}
\end{center}
\end{figure*}

\subsection{E. Extended analysis of swept angle data}

When a Fermi surface has an interlayer corrugation, as we identify for the $\gamma$ pocket (depicted in Fig.~\ref{pockets}a), angle-dependent measurements provide a means for mapping its in-plane topology~\cite{wosnitza1}.  If the degree of corrugation is sufficiently deep to yield separately resolved neck and belly frequencies in Fourier transforms~\cite{harrison1}, quantum oscillations then provide an accurate means for mapping the in-plane topology. The in-plane topology can be efficiently mapped out by sweeping $\theta$ in a constant magnetic field for many different in-plane orientations $\phi$ of the magnetic field  ${\bf H}_\|=(H\sin\theta\cos\phi,H\sin\theta\sin\phi,0)$$-$ requiring the use of a dual-axis rotator (for other examples, see Appendix K). In the present experiments on YBa$_2$Cu$_3$O$_{6+x}$ with $x=$~0.56 and 0.54, the $\gamma$ electron pocket is found to be deeply corrugated yielding spectral features between 10 and 40~\% in amplitude of the spectral weight associated with the $\alpha$ pocket. A minimal model for
fitting is given by
\begin{eqnarray}\label{waveform}
\Delta f=\Delta f_{\alpha,0}\cos\bigg(\frac{2\pi F_{\alpha,0}}{\mu_0H\cos\theta}\bigg)\exp\bigg(-\frac{\Gamma_\alpha}{\mu_0H\cos\theta}\bigg)+\Delta f_{\gamma,0}\cos\bigg(\frac{2\pi F_{\gamma,0}}{\mu_0H\cos\theta}\bigg)\exp\bigg(-\frac{\Gamma_\gamma}{\mu_0H\cos\theta}\bigg)\nonumber\\
\times{\rm J}_0\bigg(\frac{2\pi\Delta F_{\gamma,0}{\rm J}_0(k_\|c\tan\theta)}{\mu_0H\cos\theta}\bigg)
\end{eqnarray}
(similar to Eqn.~\ref{fittwo}, but without including a corrugation of $\alpha$)
that takes into consideration the superposition of $\alpha$ and $\gamma$ oscillation waveforms in Figs.~\ref{twoaxis}a and ~\ref{fits}a).
Here $\Delta f_{\alpha,0}$ and $\Delta f_{\gamma,0}$ are amplitude prefactors, $\Gamma_{\gamma}$ is a damping factor for the $\gamma$
oscillations, while the remaining parameters are defined in the main text.

Changes in the caliper radius of the $\gamma$ pocket $k_\|$ are assumed to be the dominant factor responsible for the subtle $\phi$-dependence of the waveform in Fig.~\ref{twoaxis}a of the main text and Fig.~\ref{fits}a. Fourier analysis in Fig.~\ref{fits}b of the $\theta$-swept oscillations in $1/\mu_0H\cos\theta$ yields a leading frequency  $F_\alpha\approx$~539~$\pm$~5~T similar to that obtained on sweeping the field at fixed angle, confirming the $1/\cos\theta$ angular dependence of the dominant $\alpha$ oscillations. We note that an additional angular damping mainly affecting the $\gamma$ pocket is observed on comparison of the oscillations measured as a function of angle with the oscillations measured as a function of magnetic field $-$ suggesting an interlayer decoherence factor that additionally suppresses the oscillation amplitude with increased angle.

Owing to the subtle variation of the waveform with $\phi$, parameters $\Delta f_{\alpha,0}$, $\Delta f_{\gamma,0}$, $F_{\alpha,0}$, $F_{\gamma,0}$ and $\Gamma_{\gamma}$ are obtained from fits of Eqn.~\ref{waveform} to the angular data assuming a circular cross-section (i.e $k_\|=k_{\rm F}$, where $k_{\rm F}=\sqrt{2eF_{\gamma,0}/\hbar}$ is the mean Fermi radius). The fits yield frequency parameters $F_{\alpha,0}=$~536~$\pm$~5~T and $F_{\gamma,0}=$~535~$\pm$~5~T, closely matching those determined in magnetic field sweeps. These parameters are then held constant for {\it all} $\phi$, and a single parameter fit is performed with only $k_\|$ being allowed to vary to refine the fits for each $\phi$. The resulting $\phi$-dependence of $k_\|$$-$ shown in Fig.~\ref{twoaxis}b of the main text$-$ yields a rounded-square Fermi surface cross-section for the $\gamma$ electron pocket, similar to that calculated using the helical spin-density wave model. We assume that the in-plane pocket topology accessed is an average between the bilayer split pocket topologies (such as shown in Fig.~\ref{bilayerfig}). In Fig.~\ref{fits}a, we show the $\phi$ dependence of the actual fits to Eqn.~\ref{waveform}.

\subsection*{F. Effect of bilayer splitting}
\begin{figure}[htbp]
\centering
\includegraphics*[width=.6\textwidth]{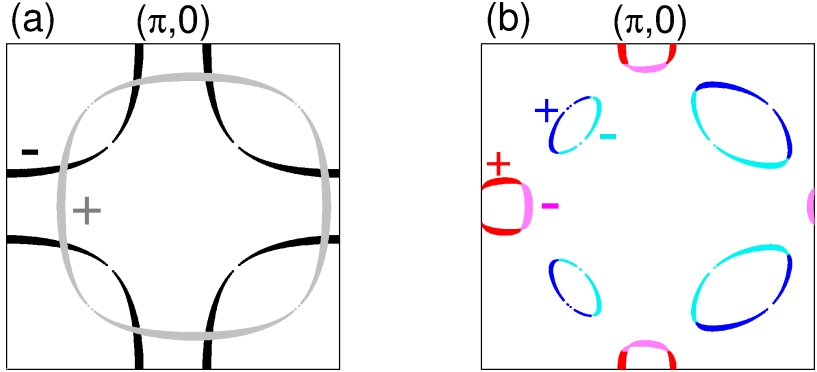}
\caption{Calculated Fermi surfaces in which the effects of bilayer splitting are included as a perturbation in r (here the effects of corrugation are not shown for clarity, but are the same as in Fig.~\ref{OKfig}). The difference in area between different bilayer-split variants of the Fermi surface is represented by the thickness of the line, where the colours (and `$+$' and `$-$' labels) indicate the sign of this difference. {\bf a} shows the translation of the unreconstructed Fermi surface where the coupling between bonding and antibonding bands is represented by different signs for the difference in area (black for `$-$' and gray for `$+$') between $\varepsilon_{\bf k}$ and $\varepsilon_{{\bf k}+{\bf Q}}$. {\bf b} shows the reconstructed Fermi surface where the difference in area (red for `$+$' and pink for `$-$' for the electron pocket, and blue for `$+$' and cyan for `$-$' for the hole pocket), can be seen to cancel for each pocket.}
\label{bilayerfig}
\end{figure}

Bilayer splitting in YBa$_2$Cu$_3$O$_{6+x}$ arises from the bilayer crystalline structure, yielding bonding and antibonding bands that lead to two large hole sections of different sizes in the paramagnetic phase (see Fig.~\ref{bilayerfig}a)~\cite{andersen1}. Here the width of the line represents the difference in area between bilayer-split bands, with `$+$' and `$-$' representing the sign of this difference.

If ordering occurs in which there is antiferromagnetic coupling between bonding and antibonding bands, the Hamiltonian factorizes into
\begin{equation}\label{matrices}
\left(
\begin{array}{cc}
\varepsilon^{\rm b}_{\bf k} & \Delta_{\rm s} \\
\Delta_{\rm s} & \varepsilon^{\rm a}_{{\bf k}+{\bf Q}} \end{array}
\right)
\hspace{0.5cm}{\rm and}\hspace{0.5cm}
\left(
\begin{array}{cc}
\varepsilon^{\rm a}_{\bf k} & \Delta_{\rm s} \\
\Delta_{\rm s} & \varepsilon^{\rm b}_{{\bf k}+{\bf Q}} \end{array}
\right).
\end{equation}
The resulting Fermi surface shown in Fig.~\ref{bilayerfig}b has two different variants$-$ the difference between them being more pronounced for the electron pocket, which is stretched along each of the $k_x$ and $k_y$ directions. The difference in area between each is represented by the thickness of the line, with the sign of the difference being positive or negative as indicated by `$+$' or `$-$.' Since the lines bounding each of the pockets contain both positive and negative segments that almost cancel, the pocket areas of the each bilayer-split variants of the reconstructed Fermi surface are only weakly split$-$ i.e. the degeneracy of the pockets is not significantly lifted by bilayer splitting when the bilayers are antiferromagnetically coupled. Consequently, different frequencies would not be expected to arise in quantum oscillation experiments~\cite{dimov1,podolsky1} from antiferromagnetically coupled bilayer split pockets.

Were the bilayers instead ferromagnetically coupled, then like bands would be coupled giving rise to significant differences in pocket areas between
each of the bilayer-split Fermi surfaces. The size of this difference is expected to significantly exceed the depth of corrugation~\cite{andersen1} and so would be observable as distinct frequencies in the quantum oscillation data, observable as well separated peaks in the Fourier transform.

\subsection{G. Inconsistency of scenario where observed multiple frequencies arise from bilayer-split pockets}
\begin{figure*}[htbp]
\centering 
\begin{center}
\includegraphics*[width=.55\textwidth]{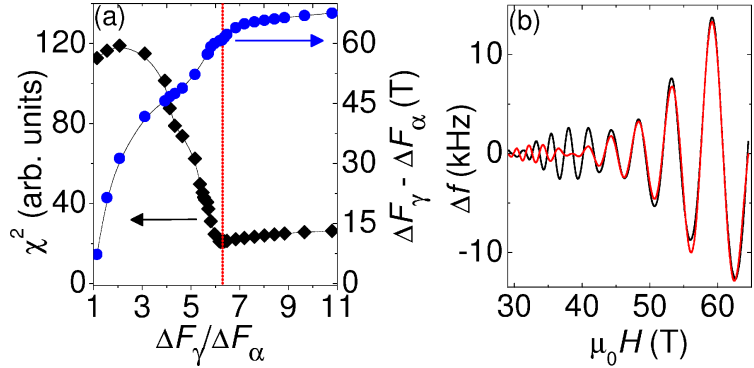}
\caption{Error analysis of unconstrained oscillation fit to two Fermi surface sections. (a) Least squares error of the unconstrained fit of Eqn.~\ref{fittwo} to the measured oscillations of the sample with oxygen concentration $x=0.56$ as a function of different ratios between the corrugation of the $\alpha$ and $\gamma$ pockets is shown on the left hand axis. The corresponding difference between the fit corrugations of the $\alpha$ and $\gamma$ pockets is shown on the right hand axis. A clear minimum is observed for $\Delta F_{\gamma,0}/\Delta F_{\alpha,0}~\approx~6.2$, yielding the best fit parameters~-~while the error is seen to rapidly increase if the corrugations of the two pockets are similar. (b) Example best fit (red line) to the measured oscillations for the scenario where $\Delta F_{\gamma,0}\approx \Delta F_{\alpha,0}$. The fit does not yield good agreement, demonstrating the need for two low frequency Fermi surface sections of significantly different warpings to explain the measured oscillations.
}
\label{fig8p5}
\end{center}
\end{figure*}

Here we show that a scenario where the observed multiple frequencies arise from two bilayer split pockets such as invoked in Ref.~\cite{audouard1} is inconsistent with the experimental data. Fig.~\ref{fig8p5} shows the least square error in an unconstrained fit of the measured oscillations to Eqn.~\ref{fittwo}. The Bessel term ${\rm J}_0(2\pi\Delta F_i/B)$ that modulates the quantum oscillation amplitude in Eqn.~\ref{fittwo}~\cite{mineev1} arises from phase smearing due to the Onsager phase of a quasi-two-dimensional Fermi surface with depth of corrugation $\Delta F_i$ that varies sinusoidally with $k_z$. In the limit $2\pi \Delta F_i/B>>1/4$, the waveform from each section $i$ is a superposition of extremal neck and belly frequencies $F_{i,{\rm neck,belly}}=F_i\mp\Delta F_i$. For corrugated cylindrical pockets with a small degree of warping, such as the $\alpha$ pocket, the difference in Onsager phase ($2\pi(2\Delta F)/\mu_0 H$) between the neck and belly extremal orbits at $\mu_0 H = 65$T is small enough ($\approx$ 2 radians for the $\alpha$ pocket) to render inadequate the parabolic approximation assumed in the Lifshitz Kosevich expression~\cite{shoenberg1}. Instead, the corrugated cylinder topology is best represented by Eqn.~\ref{fittwo} in which the amplitude of the neck and belly oscillations is constrained to be identical (i.e. $C_i$), and their relative phase difference constrained to be $\pi/2$~\cite{mineev1}.

All parameters have been allowed to vary for different fixed ratios of $\Delta F_{\gamma,0}$ and $\Delta F_{\alpha,0}$, and the resulting least squares error of the best fit obtained as a function of the corrugation ratio between the two pockets. It is evident that the minimum error appears for a significantly different corrugation of the $\alpha$ and $\gamma$ sections, with a rapidly growing error with increasingly similar corrugations of the $\alpha$ and $\gamma$ Fermi surface sections. Very similar parameters are also obtained on fitting the quantum oscillations in Ref.~\cite{audouard1} to Eqn.~\ref{fittwo} (where a sine term replaces the cosine term for magnetic oscillations), indicating consistency with two very differently warped pockets. Considering the scenario where bilayer split pockets are invoked to explain the multiple low frequencies, the warping of each of the two $\alpha$ and $\gamma$ sections would be required to be the same (we assume that the non-conducting chains in underdoped YBa$_2$Cu$_3$O$_{6+x}$ do not hybridise with bonding and anti-bonding states to create differently warped bilayer split sections as in optimally doped YBa$_2$Cu$_3$O$_{7}$~\cite{andersen1}). The least squares error is near maximum in this case, as demonstrated by the best fit for $\Delta F_{\gamma,0}/\Delta F_{\alpha,0}~\approx 1$ shown in Fig.~\ref{fig8p5}. This disagreement in fit indicates that the multiple observed low frequencies are not explained by similarly warped pockets, as would be the case for bilayer-split pockets originating from a ferromagnetic bilayer coupling.

The bilayer splitting which must occur for YBa$_2$Cu$_3$O$_{6+x}$ is not experimentally resolvable as an additional set of frequencies in the Fourier transform. Indications therefore are that the areas of the bilayer split pockets remain almost degenerate, as would occur for the antiferromagnetic coupling of bonding and antibonding states, consistent with the low energy spin excitations observed in inelastic neutron scattering experiments~\cite{fong1}.

It can readily be seen why the scenario of two equally warped Fermi surface sections does not explain the multiple low frequency oscillations. We define two pockets (labeled `1' and `2') with neck and belly frequencies $F_{1,{\rm neck}}$ and $F_{1,{\rm belly}}$ and $F_{2,{\rm neck}}$ and $F_{2,{\rm belly}}$ in terms of two median frequencies $F_1$ and $F_2$ with a depth of corrugation $\Delta F$: i.e.
\begin{eqnarray}\label{twoelectrons}
F_{1,{\rm neck/belly}}=F_1\pm\Delta F&,&F_{2,{\rm neck/belly}}=F_2\pm\Delta F.
\end{eqnarray}
where $\Delta F$ would have the same magnitude and sign for the two pockets in a ferromagnetically-coupled bilayer splitting model.

Quantum oscillations result from a superposition of two oscillatory terms
\begin{eqnarray}\label{twoterms}
\Delta f\propto\bigg[\cos\bigg(\frac{2\pi F_1}{\mu_0H\cos\theta}\bigg){\rm J}_0\bigg(\frac{2\pi\Delta F}{\mu_0H\cos\theta}\bigg)+\cos\bigg(\frac{2\pi F_2}{\mu_0H\cos\theta}\bigg){\rm J}_0\bigg(\frac{2\pi\Delta F}{\mu_0H\cos\theta}\bigg)\bigg]
\end{eqnarray}
for the two pockets, yielding the four frequencies given above. On combining them, we obtain
\begin{eqnarray}\label{singleterm}
\Delta f\propto\cos\bigg(\frac{\pi (F_1+F_2)}{\mu_0H\cos\theta}\bigg)\times\cos\bigg(\frac{\pi (F_1-F_2)}{\mu_0H\cos\theta}\bigg){\rm J}_0\bigg(\frac{2\pi\Delta F}{\mu_0H\cos\theta}\bigg)
\end{eqnarray}

Two terms in Eqn.~\ref{singleterm} will give rise to nodes: the Bessel term containing $\Delta F$ will give rise to nodes at the set of magnetic fields given by Eqn.~\ref{nodeequation} (eqn.~\ref{nodeequation} in the main text), while the cosine term containing $(F_1-F_2)/2$ will give rise to nodes
\begin{equation}\label{nodeequation2}
\mu_0H_m\cos\theta=\frac{|F_1-F_2|}{2m+1}
\end{equation}
where $m$ is an integer. Since the node terms are multiplicative, both sets of nodes would be observed giving rise to a suppression of the total waveform (as shown in the attempted fit in Fig.~\ref{fig8p5}b) at the field values given by both Eqns.~\ref{nodeequation} and \ref{nodeequation2}. The absence of such nodes in the full experimental waveform  (Fig. 2a,~\ref{angles}b of the main text) indicates that the multiple observed low frequencies are not a consequence of bilayer-split pockets. 

\subsection{H. Additional swept angle data}
\begin{figure*}[htbp]
\centering 
\includegraphics*[width=.3\textwidth]{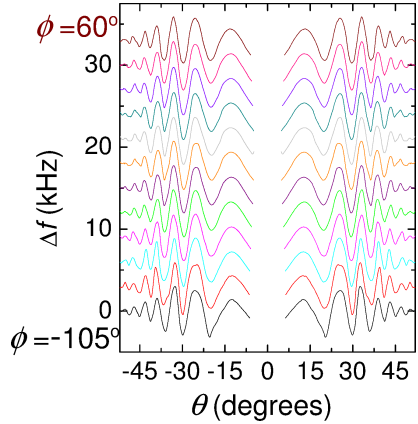}
\caption{Swept $\theta$ oscillations measured in a sample of YBa$_2$Cu$_3$O$_{6+x}$ with $x=$~0.54 at constant field $H=$~45~T and different fixed values of $\phi$ at $T\approx$~1.5~K.}
\label{fig11}
\end{figure*}

In Fig.~\ref{fig11}  we show oscillation in data measured by sweeping $\theta$ for constant $H$ at many different values of $\phi$ obtained for a single crystal sample of YBa$_2$Cu$_3$O$_{6+x}$ with $x=$~0.54. The weak $\phi$-dependence reflects the behavior observed for $x=$~0.56. 

\subsection{I. Bandstructure details}
Antiferromagnetic coupling between bilayers couples bonding to antibonding bands, yielding two slightly different variants of the Fermi surfaces shown in Figs.~\ref{OKfig}b and \ref{OKfig}c of the main text with very similar areas.

\begin{figure*}[htbp]
\centering 
\includegraphics*[width=.47\textwidth]{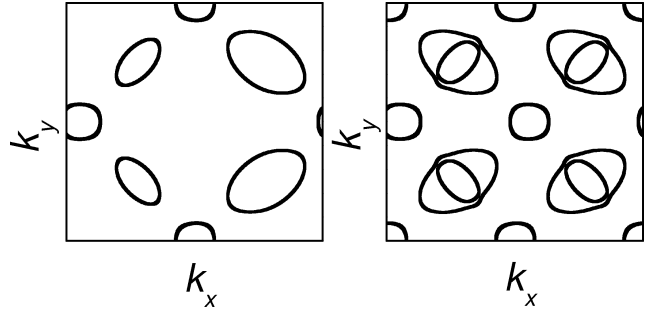}
\caption{ The left-hand panel shows the Fermi surface calculated for a helical spin-density wave model. Using a hole doping $p=$~11.7~\% and $\delta\approx\frac{1}{16}\approx$~0.06 as constraints, adjustment of a single parameter $\Delta_{\rm s}$ yields a Fermi surface with frequencies comparable to those observed$-$ i.e. $F_{\alpha,0}=$~575~T, $F_{\beta,0}=$~1583~T, and $F_{\gamma,0}=$~517~T. The largest discrepancy for the case of the $\beta$ pocket corresponds to 0.2~\% of the paramagnetic Brillouin zone area. The helical spin-density wave model used is described in Ref.~\cite{sebastian2}. On including an ortho-II potential (right hand panel), using the formalism described in Ref.~\cite{newkee}, pockets with frequencies$-$$F_{\alpha,0}=$~517~T, $F_{\beta,0}=$~1641~T, and $F_{\gamma,0}=$~515~T$-$ closer to those observed experimentally are obtained in the right-hand panel. The largest discrepancy for any one pocket now corresponds to 0.07~\% of the Brillouin zone.}
\label{fig12}
\end{figure*}

For samples with oxygen concentration close to $x\approx$~0.5, ortho-II ordering has the potential to modify the Fermi surface, depending on the strength of the relevant potential $\Delta_{\rm II}$~\cite{newkee,dimov1}. For commensurate spin ordering (i.e. $\delta=$~0), $\Delta_{\rm II}$ alone can yield pockets comparable to those measured~\cite{podolsky1}. For incommensurate ordering where $\delta\gtrsim\frac{1}{16}$, however, ortho-II makes little difference to the observed frequencies~\cite{podolsky1} (see Fig.~\ref{fig12}). In Fig.~\ref{fig12}, the ortho-II potential can been shown to provide improved agreement between the calculated Fermi surface and experiment.

\begin{figure*}[htbp]
\centering 
\includegraphics*[width=.47\textwidth]{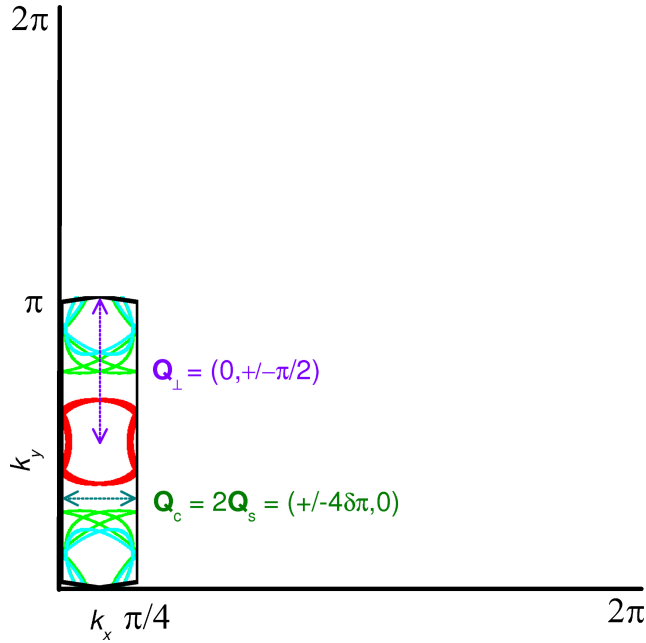}
\caption{The actual Brillouin zone for Fermi surface reconstruction by ${\bf Q}_{\rm s}=(\pi[1\pm2\delta],\pi)$, assuming $\delta=\frac{1}{16}$, illustrating the possible superlattice modulation vectors ${\bf Q}_{\rm c}$ and ${\bf Q}_\perp$ accompanying an excitonic insulator instability.}
\label{realzone}
\end{figure*}

\subsection{J. Possible ordering resulting from an excitonic insulator instability}
While the extended zone representations in Figs.~\ref{OKfig}b and c of the main text provide a convenient means of visualizing the three distinct pockets, the actual magnetic Brillouin zone depicted in Fig.~\ref{realzone} (obtained on repeatedly folding that in Fig.~\ref{OKfig}b of the main text) must be considered in order to identify possible ordering vectors associated with the excitonic insulator instability. Here, we make the simplifying assumption that $\delta=\frac{1}{16}\approx$~0.06.

One possible candidate ordering vector is ${\bf Q}_{\rm c}=2{\bf Q}_{\rm s}=(\pm4\delta\pi,0)$, corresponding to the charge modulation that can accompany a spin-density wave should it become collinear or `stripe-like' (as opposed to helical). In this case ${\bf Q}_{\rm c}$ does not introduce a new periodicity in Fig.~\ref{realzone}, but rather introduces additional couplings between $\varepsilon_{{\bf k}+n{\bf Q}_{\rm s}}$ and $\varepsilon_{{\bf k}+(n\pm2){\bf Q}_{\rm s}}$ that are not present in a purely helical spin-density wave. Such additional couplings will cause gaps to open around the edges of the magnetic Brillouin zone in Fig.~\ref{realzone}, leading to the destruction of closed pockets and the creation of open Fermi surface sheets in a manner analogous to those obtained Ref.~\cite{millis1} (where $\delta=\frac{1}{8}$ was considered). The observation of phonon broadening effects~\cite{mook2} together with the possible observation of Bragg peaks in Ref.~\cite{mook1} could be consistent with charge ordering of this type.

An alternative possibility is a form of broken translational symmetry with a characteristic vector ${\bf Q}_\perp\approx(0,\pi/2)$ that maps the electron and hole pockets onto each other as in a conventional Fermi surface nesting scenario. In this case, ${\bf Q}_\perp$ has a similar periodicity to the purely charge wavevectors identified in stripe systems~\cite{millis1} and/or scanning tunneling microscopy experiments~\cite{shen1}. While such forms of order are reported in various families of high $T_{\rm c}$ cuprates, they have not been observed in YBa$_2$Cu$_3$O$_{6+x}$. 

\begin{figure*}[ht!]
\centering 
\includegraphics*[width=.47\textwidth]{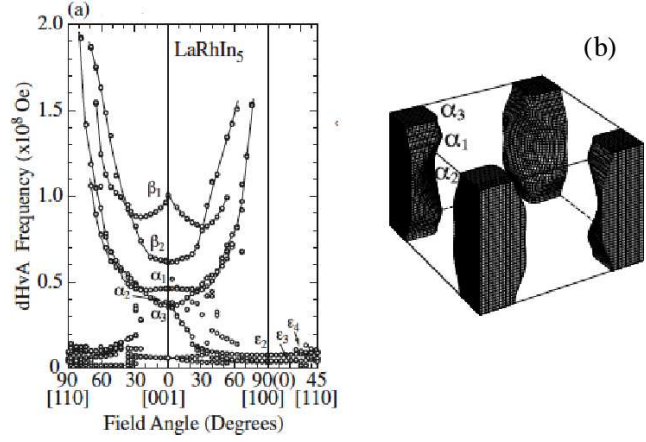}
\caption{{\bf a}, Experimentally measured angular dependence of the dHvA frequencies in LaRhIn$_5$. {\bf b} Band structure calculations of the Fermi surface topology of the $\alpha$-pockets in LaRhIn5}
\label{fig13}
\end{figure*}

\subsection{K. Rotation studies on other corrugated Fermi surface materials}
The rare-eath paramagnet LaRhIn$_5$~\cite{onuki1} and the organic superconductor $\beta$-(BEDT-TTF)$_2$IBr$_2$~\cite{wosnitza1} are two examples of strongly correlated materials for which a similar degree of corrugation enables angle-dependent quantum oscillation experiments to access information on the in-plane topology. 

\begin{figure*}[h!]
\centering 
\includegraphics*[width=.47\textwidth]{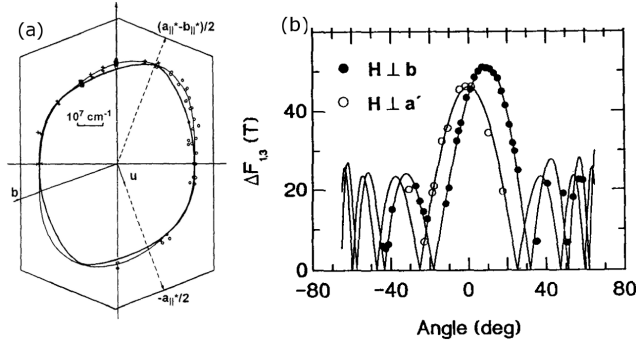}
\caption{{\bf a}, In-plane Fermi surface cross-section of $\beta$-(BEDT-TTF)$_2$IBr$_2$ from Ref.~\cite{wosnitza1}. {\bf b}, $\theta$-dependence of the difference frequency $\Delta F_{1,3}=F_{\rm belly}-F_{\rm neck}$ (also from Ref.~\cite{wosnitza1}) for two different $\phi$ angles, which 90$^\circ$ apart according to ({\bf a}).}
\label{fig10}
\end{figure*}

LaRhIn$_5$ has a section of Fermi surface (labelled $\alpha$ in Fig.~\ref{fig13}) with a proportionately similar degree of corrugation to the $\gamma$ pocket in YBa$_2$Cu$_3$O$_{6+x}$ and a round-square cross-section~\cite{onuki1}. de~Haas-van~Alphen experiments can resolve neck and belly frequencies $F_{\alpha 1,2}$ and $F_{\alpha 3}$ respectively, which become degenerate when ${\bf H}$ is rotated by an angle $\theta$ away from the $[0,0,1]$ axis, corresponding to the first Bessel zero in $F\cos\theta\approx F_0\pm\Delta F J_0(k_\|c^\prime\tan\theta)$. Owing to the rounded-square cross-section, the caliper radius is larger along $<1,1,0>$ than along $<1,0,0>$ causing the degeneracy to occur at a smaller angle $\theta$ when the field is rotated towards $[1,1,0]$ than when it is rotated towards $[1,0,0]$.

According to Ref.~\cite{wosnitza1}, $\beta$-(BEDT-TTF)$_2$IBr$_2$ has a warped Fermi surface (see Fig.~\ref{fig10}a) with an elliptical in-plane topology and triclinic crystal structure, causing the $\theta$-dependences of the difference frequency $\Delta F_{1,3}=F_{\rm belly}-F_{\rm beck}$ (see Fig.~\ref{fig10}b) to be different for the two different $\phi$ angles considered (which are 90$^\circ$ apart). The existence of only a single section of Fermi surface $\beta$-(BEDT-TTF)$_2$IBr$_2$ and the very low transition temperature of this organic superconductor enables multiple nodes to be directly observed. 

\section{Acknowledgements}
This work is supported by US Department of Energy, the National Science Foundation (incl. Grant No. PHY05-51164), the State of Florida, the Royal Society, Trinity College (University of Cambridge), the EPSRC, and the BES program `Science in 100 T'. The authors thank P. B. Littlewood for theoretical input, B. Ramshaw for discussions, and M. Gordon, A. Paris, D. Rickel, D. Roybal, and C. Swenson for technical assistance.

\end{document}